\documentclass[twocolumn, journal]{IEEEtran}
\IEEEoverridecommandlockouts

\usepackage{color}
\usepackage{bm}
\usepackage{graphicx}
\usepackage{amsmath}
\usepackage{amssymb}
\usepackage{algorithm}
\usepackage{algorithmic}
\usepackage{amsmath}
\usepackage{multirow}
\usepackage{booktabs}
\usepackage{array}
\usepackage{amsthm}
\usepackage{lipsum}
\usepackage{enumerate}
\usepackage{stfloats}
\usepackage{subfigure}
\usepackage{cases}
\usepackage{cite}
\usepackage{makecell}
\usepackage{soul}
\usepackage{diagbox}

\newtheorem{remark}{Remark}

\title{Beyond Diagonal Reconfigurable Intelligent Surfaces: From Transmitting and Reflecting Modes to Single-, Group-, and Fully-Connected Architectures}

\author{Hongyu Li,~\IEEEmembership{Student Member,~IEEE}, Shanpu Shen,~\IEEEmembership{Member,~IEEE}, and Bruno Clerckx,~\IEEEmembership{Fellow,~IEEE}
\thanks{Manuscript received; \textit{(Corresponding author: Shanpu Shen).}}
\thanks{H. Li and B. Clerckx are with the Department of Electrical and Electronic Engineering, Imperial College London, London SW7 2AZ, U.K. (email:\{c.li21,b.clerckx\}@imperial.ac.uk).}\\
\thanks{S. Shen is with the Department of Electronic and Computer Engineering, The Hong Kong University of Science and Technology, Clear Water Bay, Kowloon, Hong Kong (email: sshenaa@connect.ust.hk).}}

\begin{document}

\maketitle
\thispagestyle{empty}

\vspace{-1.5 cm}

\begin{abstract}

Reconfigurable intelligent surfaces (RISs) are envisioned as a promising technology for future wireless communications. 
With various hardware realizations, RISs can work under different modes (reflective/transmissive/hybrid) or have different architectures (single/group/fully-connected). 
However, most existing research focused on single-connected reflective RISs, mathematically characterized by diagonal phase shift matrices, while there is a lack of a comprehensive study for RISs unifying different modes/architectures. 
In this paper, we solve this issue by analyzing and proposing a general RIS-aided communication model. Specifically, we establish an RIS model not limited to diagonal phase shift matrices, a novel branch referred to as beyond diagonal RIS (BD-RIS), unifying modes and architectures. 
With the proposed model, we develop efficient algorithms to jointly design transmit precoder and BD-RIS matrix to maximize the sum-rate for RIS-aided systems. 
We also provide simulation results to compare the performance of BD-RISs with different modes/architectures. 
Simulation results show that under the same mode, fully- and group-connected RIS can effectively increase the sum-rate performance compared with single-connected RIS, and that hybrid RIS outperforms reflective/transmissive RIS with the same architecture.

\end{abstract}


\section{Introduction}

Reconfigurable intelligent surfaces (RISs), which can build controllable radio environments and improve communication quality in a cost-effective manner \cite{DiRenzo2020,SGong2019,K-KWong}, have recently been regarded as a revolutionary technology for wireless communication research. 
An RIS is a two-dimensional planar surface which consists of numerous nearly passive elements with ultra-low power consumption. Each element has a multi-layer structure composed of rectangle patches and tuneable devices, e.g., positive intrinsic negative (PIN) diodes.
Each PIN diode of each element can be independently switched between ``ON'' and ``OFF'' states, thereby generating different responses (i.e., phase shifts and amplitudes) for incident signals \cite{QWu2019}. Thus, RISs can be flexibly deployed into various wireless communication systems to modify the propagation environment without much power consumption.

There are various tunable surface designs for RIS realizations. On one hand, when each element of the RIS has different hardware realizations (i.e., the number of layers, size, thickness, the number and distributions of PIN diodes), the RIS can work under three \textit{modes}: transmission, reflection, and hybrid transmission and reflection \cite{JXuSTAR}, \cite{SZhang}.
As for the reflective/transmissive RIS, incident signals are either reflected from or transmitted through the RIS towards the same/opposite side as the transmitter. For the hybrid transmissive and reflective RIS, incident signals are split into two parts and can simultaneously arrive at both sides of the RIS.
On the other hand, RISs with different architectures have recently been modeled and designed by using scattering parameter network analysis \cite{SShen}. According to the circuit topology among different RIS elements, RIS can be classified into three kinds of \textit{architectures}:
single-, group-, and fully-connected architectures.
The single-connected architecture, in which each element is not connected with each other, has been widely considered in existing research. When all/part of RIS elements are connected with each other, fully/group-connected architectures can be implemented to further enhance the RIS performance.

The conventional RIS, which essentially has the single-connected architecture and works under the reflective mode yielding mathematically a \textit{diagonal phase shift matrix}, is most commonly considered in previous research \cite{YHan,RZhang,YLiu2021,BDi,QWu,JXu,ZLi,BZhengMulti-IRS,WMei,Kris,YangZhao,BZheng2020,CLiu,SAbeywickrama,WCai,HLi,WYang}. 
Beamforming designs for single RIS-aided wireless communication systems were developed using different metrics, e.g., power minimization, rate maximization, max-min fairness \cite{YHan,RZhang,YLiu2021}. Considering the high hardware complexity and cost of realizing RISs with infinite/high-resolution phase shifts, research on practical finite/low-resolution cases was also developed \cite{BDi,QWu,JXu}.
To improve the system performance, some researchers studied the deployment of multiple RISs and coordination problems \cite{ZLi,BZhengMulti-IRS,WMei}. 
In addition, RIS has also been investigated to provide performance enhancement in various scenarios, e.g., Wireless Power Transfter (WPT) \cite{Kris}, Simultaneous Wireless Information and Power Transfer (SWIPT) \cite{YangZhao}. 
Different channel estimation algorithms for RIS-aided systems were proposed based on either traditional communication theories \cite{BZheng2020} or machine learning approaches \cite{CLiu}.
While the aforementioned research on reflective single-connected RIS \cite{YHan,RZhang,YLiu2021,BDi,QWu,JXu,ZLi,BZhengMulti-IRS,WMei,Kris,YangZhao,BZheng2020,CLiu} assumes an ideal RIS reflection model, practical RIS model analysis has been proposed for both narrowband \cite{SAbeywickrama} and wideband systems \cite{WCai,HLi,WYang}.

Limited work has been carried out for RIS models \textit{beyond diagonal phase shift matrices}. 
\cite{SShen} first proposed to analyze the modeling of RIS based on the scattering parameter network and categorized the RIS into single-, group-, and fully-connected RIS architectures. Interestingly, in contrast to conventional RIS that relies on diagonal phase shift (scattering) matrix to only adjust the phases of the impinging waves, the group- and fully-connected RIS rely on more controllable scattering matrix, not limited to diagonal matrix, which enables them to adjust not only the phases but also the magnitudes of the impinging waves, therefore providing additional performance enhancements over conventional RIS. 
Following this work, the authors in \cite{Matteo} further considered designs of single/group/fully-connected RIS with discrete values. 
Results suggest that while four resolution bits are needed in single-connected RIS, only a single resolution bit is sufficient in fully-connected RIS, simplifying significantly the future development of these promising RIS architectures. 
Motivated by those benefits of RIS beyond diagonal phase matrices, authors in \cite{QLi} branched out to propose another novel RIS architecture with non-diagonal phase shift matrices, which resulted in a higher rate compared to conventional single-connected cases.
Recently, some researchers have started to focus on RISs working under the hybrid mode \cite{JXuSTAR,SZhang,HZhang,JXuHybridRIS}, whose mathematical model is also beyond diagonal phase shift matrix. Specifically, \textit{Xu et al.} \cite{JXuHybridRIS} introduced the concept of hybrid transmissive and reflective RIS, which was also referred to as simultaneously transmitting and reflecting RIS (STAR-RIS) or intelligent omni-surface (IOS) \cite{JXuSTAR,HZhang}, generated hardware and channel models, and analyzed the achievable diversity gain.

\begin{figure}
    \centering
    \includegraphics[width=0.48\textwidth]{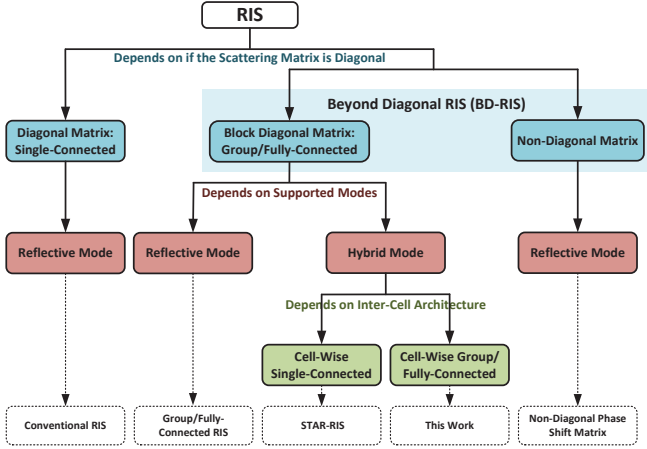}
    \caption{RIS classification tree.}
    \label{fig:RIS_tree} \vspace{-0.5 cm}
\end{figure}

To establish a clear classification of RIS, we start from the mathematical characteristic of the scattering matrix of RIS, categorizing RIS into diagonal RIS and beyond diagonal RIS (BD-RIS). Namely, the scattering matrix is (not) restricted to be diagonal, as illustrated in Fig. \ref{fig:RIS_tree}. 
Based on the RIS tree in Fig. \ref{fig:RIS_tree}, we classify the aforementioned studies as diagonal RIS under the reflective mode \cite{YHan,RZhang,YLiu2021,BDi,QWu,JXu,ZLi,BZhengMulti-IRS,WMei,Kris,YangZhao,BZheng2020,CLiu,SAbeywickrama,WCai,HLi,WYang} and BD-RIS, which includes block-diagonal RIS (group/fully-connected RIS) under the reflective mode \cite{SShen,Matteo} and under the hybrid mode (STAR-RIS) \cite{JXuSTAR,SZhang,HZhang,JXuHybridRIS}, and non-diagonal RIS under the reflective mode \cite{QLi}.
In this paper, we provide a thorough analysis of BD-RIS and extend new branches of the RIS tree. 
The main contributions of this paper are summarized as follows:

\textit{First,} we analyze and propose a general BD-RIS-aided communication model using microwave network theory. 
The proposed model is derived by characterizing an RIS as multiple antennas connected to a group-connected multi-port reconfigurable impedance network. 
In conventional (diagonal) RIS, there are no connections among different ports of the multi-port reconfigurable impedance network, which is unable to support different modes. 
However, by establishing connections among different ports, we can realize more flexible modes, e.g., the hybrid mode as shown in Fig. \ref{fig:RIS_tree}.  

\textit{Second,} we prove that STAR-RIS \cite{JXuSTAR,SZhang,HZhang,JXuHybridRIS} is mathematically a special case of block diagonal RIS and physically based on group-connected reconfigurable impedance network when each two ports are connected to each other. Specifically, we introduce the concept of ``cell'', which includes each 2 antenna ports connected to each other via reconfigurable impedance components. Then STAR-RIS belongs to the branch ``cell-wise single-connected'' in Fig. \ref{fig:RIS_tree}, which indicates there are no connections among different cells.

\begin{table*}[t]
    \caption{Nine Cases of Beyond Diagonal RIS}
    \centering 
        \begin{tabular}{|c|c|c|c|}
            \hline 
            \diagbox{Modes}{Architectures} & Cell-Wise Single-Connected & Cell-Wise Group-Connected & Cell-Wise Fully-Connected\tabularnewline
            \hline 
            Reflective & $|\phi_{\mathrm{r},m}|=1,\forall m \in \mathcal{M}$ & $\mathbf{\Phi}_{\mathrm{r},g}^{H}\mathbf{\Phi}_{\mathrm{r},g}=\mathbf{I}_{\bar{M}},\forall g\in\mathcal{G}$ & $\mathbf{\Phi}_{\mathrm{r}}^{H}\mathbf{\Phi}_{\mathrm{r}}=\mathbf{I}_{M}$\tabularnewline
            \hline 
            Transmissive & $|\phi_{\mathrm{t},m}|=1,\forall m \in \mathcal{M}$ & $\mathbf{\Phi}_{\mathrm{t},g}^{H}\mathbf{\Phi}_{\mathrm{t},g}=\mathbf{I}_{\bar{M}},\forall g\in\mathcal{G}$ & $\mathbf{\Phi}_{\mathrm{t}}^{H}\mathbf{\Phi}_{\mathrm{t}}=\mathbf{I}_{M}$\tabularnewline
            \hline 
            Hybrid & $|\phi_{\mathrm{r},m}|^{2}+|\phi_{\mathrm{t},m}|^{2}=1,\forall m \in \mathcal{M}$ & $\mathbf{\Phi}_{\mathrm{r},g}^{H}\mathbf{\Phi}_{\mathrm{r},g}+\mathbf{\Phi}_{\mathrm{t},g}^{H}\mathbf{\Phi}_{\mathrm{t},g}=\mathbf{I}_{\bar{M}},\forall g\in\mathcal{G}$ & $\mathbf{\Phi}_{\mathrm{r}}^{H}\mathbf{\Phi}_{\mathrm{r}}+\mathbf{\Phi}_{\mathrm{t}}^{H}\mathbf{\Phi}_{\mathrm{t}}=\mathbf{I}_{M}$\tabularnewline
            \hline 
        \end{tabular}\label{Table_1}\\
\end{table*}

\textit{Third,} we go beyond cell-wise single-connected BD-RIS (STAR-RIS) and create a new branch,  namely ``cell-wise group/fully-connected'' as shown in Fig. \ref{fig:RIS_tree}, which means part of/all the cells are connected to each other.
We investigate three modes (reflective, transmissive, and hybrid) and three architectures (cell-wise single-, group-, and fully-connected), in total nine cases as summarized in Table \ref{Table_1}. This is the first paper to characterize RIS by taking nine different modes/architectures into consideration. The nine cases of the proposed BD-RIS model contain the modeling of conventional RIS \cite{YHan,RZhang,YLiu2021,BDi,QWu,JXu,ZLi,BZhengMulti-IRS,WMei,Kris,YangZhao,BZheng2020,CLiu} and group/fully-connected RIS \cite{SShen,Matteo} (the first row in Table \ref{Table_1}), and STAR-RIS \cite{JXuSTAR,SZhang,HZhang,JXuHybridRIS} (the third row and the first column of Table \ref{Table_1}).

\textit{Fourth,} with the proposed BD-RIS model, we consider the joint transmit precoder and BD-RIS matrix design to maximize the sum-rate for an RIS-aided multiuser multiple input single output (MU-MISO) system. 
Specifically, we first propose a general solution suitable for nine modes/architectures. Then for cell-wise single-connected case, we propose an efficient solution, which has similar performance, but lower computational complexity compared to the general one. 
We also provide initialization, convergence, and complexity analysis of the proposed design. 

\textit{Fifth,} we provide simulation results to evaluate the performance of the proposed design. We compare the sum-rate performance when the BD-RIS is under nine different modes/architectures.
Simulation results with specific parameter settings show that under Rayleigh fading conditions, cell-wise fully- and group-connected hybrid BD-RISs can achieve 75\% and 37\% higher sum-rate than cell-wise single-connected hybrid ones. In the meanwhile, the sum-rate achieved by cell-wise fully-connected hybrid BD-RISs over that by cell-wise fully-connected reflective/transmissive ones can be around 20\% for Rician fading channels.

\textit{Organization:} Section II introduces a BD-RIS-aided communication model. Then Section III presents a generalized RIS model based on architecture/mode analysis and design. With the proposed model, Section IV formulates an optimization problem and illustrates solutions for the formulated problem. 
Section V evaluates the performance of the proposed design. 
Finally, Section VI provides conclusions and future work.

\textit{Notations}:
Boldface lower-case and upper-case letters indicate column vectors and matrices, respectively.
$\mathbb{C}$ and $\mathbb{R}$ denote the set of complex and real numbers, respectively.
$\mathbb{E}\{\cdot\}$ represents statistical expectation.
$(\cdot)^\ast$, $(\cdot)^T$, $(\cdot)^H$, and $(\cdot)^{-1}$ denote the conjugate, transpose, conjugate-transpose operations, and inversion, respectively.
$\Re \{ \cdot \}$ denotes the real part of a complex number.
$\mathbf{I}_L$ indicates an $L \times L$ identity matrix.
$\mathbf{0}$ denotes an all-zero matrix.
$\| \mathbf{A} \|_F$ denotes the Frobenius norm of matrix $\mathbf{A}$.
$|a|$ denotes the norm of variable $a$.
$|\mathcal{A}|$ denotes the size of set $\mathcal{A}$.
$\mathsf{diag}(\cdot)$ denotes a diagonal matrix.
$\mathsf{blkdiag}(\cdot)$ denotes a block matrix such that the main-diagonal blocks are matrices and all off-diagonal blocks are zero matrices.
$\jmath = \sqrt{-1}$ denotes imaginary unit.
$\mathsf{Tr}(\cdot)$ denotes the summation of diagonal elements of a matrix. 
Finally, $[\mathbf{A}]_{i,:}$, $[\mathbf{A}]_{:,j}$, and $[\mathbf{A}]_{i,j}$ denote the $i$-th row, the $j$-th column, and the $(i,j)$-th element of matrix $\mathbf{A}$, respectively.

\section{BD-RIS-Aided Communication Model}

\begin{figure}
    \centering
    \includegraphics[height=1.7 in]{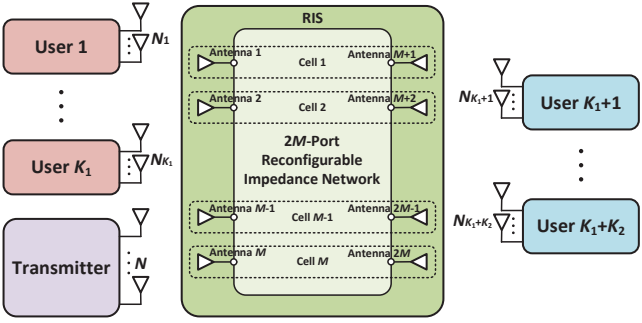}
    \caption{Diagram of an $M$-cell RIS-aided communication system.}\label{fig:RIS_NA}
\end{figure}

Consider an RIS-aided wireless communication system, where an $N$-antenna transmitter indexed by $\mathcal{N}=\{1,\ldots,N\}$ serves $K$ multi-antenna users with the assistance of an RIS. Each user has $N_{k}$ antennas, $\forall k\in\mathcal{K}=\{1,\ldots,K\}$. 
In general, we consider an $M$-cell\footnote{Here we use ``cell'' instead of ``element'' since in existing works each RIS element contains one antenna, while in this work each cell has two antennas. The ``cell'' is illustrated in Fig. \ref{fig:RIS_NA}.} RIS indexed by $\mathcal{M}=\{1,\ldots,M\}$ where the $m$th cell consists of antenna $m$ and antenna $m+M$. Therefore, in this $M$-cell RIS, there are 2$M$ antennas connected to a 2$M$-port reconfigurable impedance network, as shown in Fig. \ref{fig:RIS_NA}. 
According to \cite{SShen}, assuming that the multiple antennas at the transmitter/receiver/RIS are perfectly matched with no mutual coupling, the overall channel matrix between the transmitter
and each user, $\mathbf{H}_{\mathrm{all},k}\in\mathbb{C}^{N_{k}\times N}$, $k\in\mathcal{K}$, can be expressed as 
\begin{equation}
\mathbf{H}_{\mathrm{all},k}=\mathbf{H}_{\mathrm{d},k}+\overline{\mathbf{H}}_{k}\mathbf{\Phi}\overline{\mathbf{G}},\forall k\in\mathcal{K},\label{eq:overall_channel}
\end{equation}
where $\mathbf{H}_{\mathrm{d},k}\in\mathbb{C}^{N_{k}\times N}$, $\overline{\mathbf{H}}_{k}\in\mathbb{C}^{N_{k}\times2M}$, $\forall k\in\mathcal{K}$, and $\overline{\mathbf{G}}\in\mathbb{C}^{2M\times N}$ denote channel matrices from the transmitter to user $k$, from the RIS to user $k$, and from the transmitter to the RIS, respectively.
$\mathbf{\Phi}\in\mathbb{C}^{2M\times2M}$ denotes the scattering matrix of the $2M$-port reconfigurable impedance network \cite{SShen}.
According to microwave network theory \cite{SShen,DMPozar}, $\mathbf{\Phi}$ should satisfy $\mathbf{\Phi}^{H}\mathbf{\Phi}\preceq\mathbf{I}_{2M}$.
Particularly, when the $2M$-port reconfigurable impedance network is lossless, we have$\mathbf{\Phi}^{H}\mathbf{\Phi}=\mathbf{I}_{2M}$\footnote{Here we should explain that matrix $\mathbf{\Phi}$ is a scattering matrix, which provides an integrated description of a network as seen at its $2M$ ports. The scattering matrix associates the voltage of incident waves with reflected waves from the ports. When the multi-port network is lossless, no power can be delivered to the network so that the incident and reflected power are equal to each other. In this case, it can be derived that the scattering matrix satisfies a unitary constraint \cite{DMPozar}.}. It is worth noting that the structure of the matrix $\mathbf{\Phi}$ is determined by the circuit topology of the $2M$-port reconfigurable impedance network. Therefore, different from traditional RIS model with a diagonal matrix, the matrix $\mathbf{\Phi}$ is not restricted to be diagonal, which is referred to as BD-RIS.

\begin{figure}
    \centering 
    \includegraphics[width=0.45\textwidth]{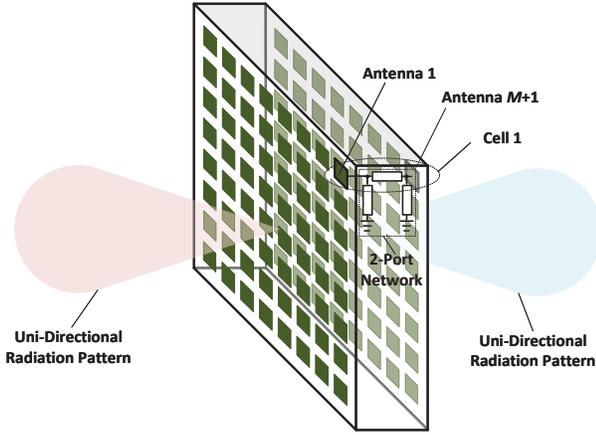}
    \caption{An $M$-cell BD-RIS with 2$M$ back to back placed uni-directional antennas.}
    \label{fig:assumption} 
\end{figure}

To show that existing RIS-aided communication model \cite{JXuSTAR,SZhang,HZhang,JXuHybridRIS} is a special case of our proposed model, we first make the following assumptions:
\begin{itemize}
\item[A1:] Each antenna has a uni-directional radiation pattern\footnote{In practice the uni-directional radiation pattern can be achieved by using microstrip patch antennas or by adding a reflecting ground plane.}.
\item[A2:] In each cell, the two antennas are back to back placed so that each antenna covers half space as shown in Fig. \ref{fig:assumption}.
\end{itemize}
With these assumptions, we divide the $M$-cell BD-RIS into 2 sectors, where the first sector consists of antennas 1-$M$ and the other sector consists of antennas $M$+1-2$M$, and thus the $M$-cell RIS partitions the whole space into two sides, which are respectively covered by the two sectors. Accordingly, we can partition the overall channel matrix (\ref{eq:overall_channel}) as 
\begin{equation}
\begin{aligned}\mathbf{H}_{\mathrm{all},k}= & \mathbf{H}_{\mathrm{d},k}+\left[\overline{\mathbf{H}}_{k,1}~\overline{\mathbf{H}}_{k,2}\right]\left[\begin{array}{cc}
\mathbf{\Phi}_{1,1} & \mathbf{\Phi}_{1,2}\\
\mathbf{\Phi}_{2,1} & \mathbf{\Phi}_{2,2}
\end{array}\right]\left[\begin{array}{c}
\overline{\mathbf{G}}_{1}\\
\overline{\mathbf{G}}_{2}
\end{array}\right]\\
= & \mathbf{H}_{\mathrm{d},k}+\overline{\mathbf{H}}_{k,1}\mathbf{\Phi}_{1,1}\overline{\mathbf{G}}_{1}+\overline{\mathbf{H}}_{k,2}\mathbf{\Phi}_{2,1}\overline{\mathbf{G}}_{1}\\
&+\overline{\mathbf{H}}_{k,1}\mathbf{\Phi}_{1,2}\overline{\mathbf{G}}_{2}+\overline{\mathbf{H}}_{k,2}\mathbf{\Phi}_{2,2}\overline{\mathbf{G}}_{2},\forall k\in\mathcal{K},
\end{aligned}
\label{eq:overall_channel1}
\end{equation}
where $\overline{\mathbf{H}}_{k,i}=[\overline{\mathbf{H}}_{k}]_{:,(i-1)M+1:iM}\in\mathbb{C}^{N_{k}\times M}$, $\forall k\in\mathcal{K}$, and $\overline{\mathbf{G}}_{i}=[\overline{\mathbf{G}}]_{(i-1)M+1:iM,:}\in\mathbb{C}^{M\times N}$, $\forall i\in\{1,2\}$ are channels between sector $i$ of the RIS and user $k$, and between the transmitter and sector $i$ of the RIS, respectively. $\mathbf{\Phi}_{i,j}=[\mathbf{\Phi}]_{(i-1)M+1:iM,(j-1)M+1:jM}\in\mathbb{C}^{M\times M}$,
$\forall i,j\in\{1,2\}$.

\begin{figure}
    \centering
    \includegraphics[width=0.45\textwidth]{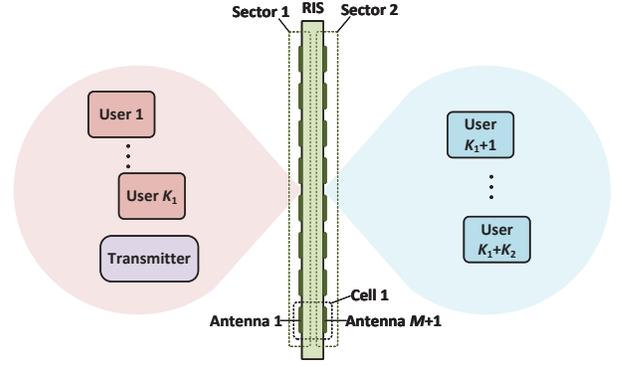}
    \caption{Top view for the locations of the $M$-cell BD-RIS partitioning the whole space into two sides, the transmitter, and multiple users.}
    \label{fig:assumption_TV}
\end{figure}

Without loss of generality, we assume that the transmitter and $K_{1}$ users, indexed by $\mathcal{K}_{1}=\{1,\ldots,K_{1}\}$, $0<K_{1}<K$, are located at one side of the RIS (covered by sector 1), and $K_{2}=K-K_{1}$ users, indexed by $\mathcal{K}_{2}=\{K_{1}+1,\ldots,K\}$, are located at the other side of the RIS (covered by sector 2). To facilitate understanding, we illustrate the locations of the transmitter, RIS, and users from a top view in Fig. \ref{fig:assumption_TV}. Following assumptions A1 and A2, we can deduce the following two corollaries:
\begin{itemize}
\item[C1:] The channel from the transmitter to sector 2 of the BD-RIS is zero, that is $\overline{\mathbf{G}}_{2}=\mathbf{0}$, as the transmitter is not covered by the uni-directional radiation pattern of sector 2.
\item[C2:] The channel from sector $i$ of the BD-RIS to the user belonging to $\mathcal{K}_{j}$
is zero $\forall i\ne j$, that is $\overline{\mathbf{H}}_{k,i}=\mathbf{0}$, $\forall k\in\mathcal{K}_{j}$, $\forall i\ne j$, as the user belonging to $\mathcal{K}_{j}$ is not covered by the uni-directional radiation pattern of sector $i$.
\end{itemize}
Leveraging corollaries C1 and C2, we can simplify the channel matrix (\ref{eq:overall_channel1}) as 
\begin{equation}
\mathbf{H}_{\mathrm{all},k}=\left\{ \begin{array}{cc}
\mathbf{H}_{\mathrm{d},k}+\overline{\mathbf{H}}_{k,1}\mathbf{\Phi}_{1,1}\overline{\mathbf{G}}_{1}, & k\in\mathcal{K}_{1},\\
\mathbf{H}_{\mathrm{d},k}+\overline{\mathbf{H}}_{k,2}\mathbf{\Phi}_{2,1}\overline{\mathbf{G}}_{1}, & k\in\mathcal{K}_{2}.
\end{array}\right.\label{eq:simplified_overall_channel}
\end{equation}
We denote users within the coverage of sector 1 of the BD-RIS as reflective users, indexed by $\mathcal{K}_{\mathrm{r}}=\mathcal{K}_{1}$, $K_{\mathrm{r}}=K_{1}$.
Similarly, we denote users within the coverage of sector 2 as transmissive users, indexed by $\mathcal{K}_{\mathrm{t}}=\mathcal{K}_{2}$, $K_{\mathrm{t}}=K_{2}=K-K_{\mathrm{r}}$.
Then, using auxiliary notations, $\mathbf{H}_{k}=\overline{\mathbf{H}}_{k,i}$,
$\forall k\in\mathcal{K}_{i}$, $\forall i\in\{1,2\}$, $\mathbf{\Phi}_{\mathrm{r}}=\mathbf{\Phi}_{1,1}$, $\mathbf{\Phi}_{\mathrm{t}}=\mathbf{\Phi}_{2,1}$, and $\mathbf{G}=\overline{\mathbf{G}}_{1}$, we can rewrite (\ref{eq:simplified_overall_channel}) as 
\begin{equation}
\mathbf{H}_{\mathrm{all},k}=\left\{ \begin{array}{cc}
\mathbf{H}_{\mathrm{d},k}+\mathbf{H}_{k}\mathbf{\Phi}_{\mathrm{r}}\mathbf{G}, & k\in\mathcal{K}_{\mathrm{r}},\\
\mathbf{H}_{\mathrm{d},k}+\mathbf{H}_{k}\mathbf{\Phi}_{\mathrm{t}}\mathbf{G}, & k\in\mathcal{K}_{\mathrm{t}}.
\end{array}\right.\label{eq:simplified_overall_channel1}
\end{equation}
Given that $\mathbf{\Phi}^{H}\mathbf{\Phi}=\mathbf{I}_{2M}$, $\mathbf{\Phi}_{\mathrm{t}}$ and $\mathbf{\Phi}_{\mathrm{r}}$ should satisfy the following constraint: 
\begin{equation}
\mathbf{\Phi}_{\mathrm{r}}^{H}\mathbf{\Phi}_{\mathrm{r}}+\mathbf{\Phi}_{\mathrm{t}}^{H}\mathbf{\Phi}_{\mathrm{t}}=\mathbf{I}_{M}.\label{eq:ris_constraint}
\end{equation}

Based on the above analysis, we have verified that the communication model utilized in STAR-RIS \cite{JXuSTAR,SZhang,HZhang,JXuHybridRIS} is a special instance of our models (\ref{eq:simplified_overall_channel1}) and (\ref{eq:ris_constraint}).
In the following section, we will go beyond STAR-RIS and investigate different cases of constraint (\ref{eq:ris_constraint}).

\section{Architecture/Mode Analysis and Design}

\subsection{Architecture Analysis and Design}

In \cite{SShen}, three kinds of RIS architectures which have different RIS element circuit topologies have been proposed. In this subsection, we generalize this concept by analyzing and designing the $M$-cell BD-RIS architecture\footnote{The following single/fully/group-connected architectures refer to the inter-cell circuit topologies. The 2 antennas within the same cell are connected to a 2-port fully-connected reconfigurable impedance network.} with different cell circuit topologies as detailed below.

\subsubsection{Cell-Wise Single-Connected (CW-SC) Architecture}
\begin{figure}
    \centering
    \includegraphics[height=3.2 in]{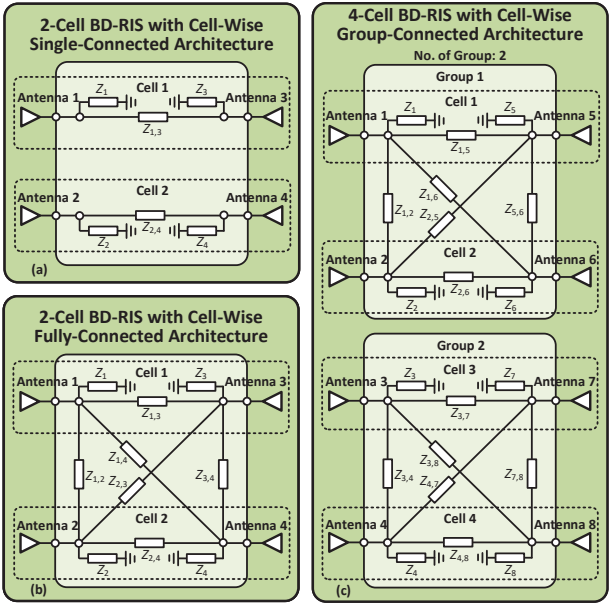}
    \caption{Examples of (a) CW-SC BD-RIS, (b) CW-FC BD-RIS, and (c) CW-GC BD-RIS.}\label{fig:architecture_eg} 
\end{figure}

In this architecture, cells of the BD-RIS are not connected to each other. 
Fig. \ref{fig:architecture_eg}(a) provides a simple example of CW-SC BD-RIS with 2 cells. Thus, matrices $\mathbf{\Phi}_{\mathrm{r}}$, $\mathbf{\Phi}_{\mathrm{t}}$ are all diagonal, which are given by
\begin{equation}
    \label{eq:Single_connected}
    \begin{aligned}
    \mathbf{\Phi}_{\mathrm{r}}&=\mathsf{diag}(\phi_{\mathrm{r},1},\ldots,\phi_{\mathrm{r},M}),~~
    \mathbf{\Phi}_{\mathrm{t}}&=\mathsf{diag}(\phi_{\mathrm{t},1},\ldots,\phi_{\mathrm{t},M}),
    \end{aligned}
\end{equation}
where the entries $\phi_{\mathrm{t/r},m},\forall m\in\mathcal{M}$, satisfy the constraints
\begin{equation}
|\phi_{\mathrm{r},m}|^{2}+|\phi_{\mathrm{t},m}|^{2}=1,\forall m\in\mathcal{M}.
\end{equation}
The BD-RIS with a CW-SC architecture is essentially the STAR-RIS \cite{JXuSTAR,SZhang,HZhang,JXuHybridRIS}. 

\subsubsection{Cell-Wise Fully-Connected (CW-FC) Architecture}
We propose and design a generalized RIS architecture referred to as CW-FC architecture where all cells of the BD-RIS are connected to each other through reconfigurable impedance components. 
In Fig. \ref{fig:architecture_eg}(b), we give an example of CW-FC BD-RIS with 2 cells.
Thus, $\mathbf{\Phi}_{\mathrm{r}}$, $\mathbf{\Phi}_{\mathrm{t}}$ are all full matrices satisfying the constraint 
\begin{equation}
\mathbf{\Phi}_{\mathrm{r}}^{H}\mathbf{\Phi}_{\mathrm{r}}+\mathbf{\Phi}_{\mathrm{t}}^{H}\mathbf{\Phi}_{\mathrm{t}}=\mathbf{I}_{M}.\label{eq:Fully_connected}
\end{equation}
CW-FC architecture can achieve the best performance due to the most general constraint (\ref{eq:Fully_connected}).

\subsubsection{Cell-Wise Group-Connected (CW-GC) Architecture}

It is obvious that when the number of cells $M$ gets larger, the circuit topology of the CW-FC BD-RIS will become extremely complicated. To achieve a good trade-off between the RIS performance and circuit complexity, we propose and design a generalized BD-RIS architecture referred to as CW-GC architecture. In the CW-GC BD-RIS, the $M$ cells are divided into $G$ groups indexed by $\mathcal{G}=\{1,\ldots,G\}$ with each group of cells utilizing the CW-FC architecture.
For simplicity, here we assume that all groups have the same size $\bar{M}=M/G$, and we define $\mathcal{G}_{g}\triangleq\{(g-1)\bar{M}+1,\ldots,g\bar{M}\}$ as the set of cell indexes for group $g$.
An example of a 4-cell BD-RIS with CW-GC architecture having 2 groups is illustrated in Fig. \ref{fig:architecture_eg}(c).
It should be noted that there are various grouping strategies while exploring different
kinds of grouping strategies is left as future work. Thus, $\mathbf{\Phi}_{\mathrm{r}}$, $\mathbf{\Phi}_{\mathrm{t}}$ are block diagonal matrices, which are given by 
\begin{equation}
\begin{aligned}\mathbf{\Phi}_{\mathrm{r}}=\mathsf{blkdiag}(\mathbf{\Phi}_{\mathrm{r},1},\ldots,\mathbf{\Phi}_{\mathrm{r},G}),\\
\mathbf{\Phi}_{\mathrm{t}}=\mathsf{blkdiag}(\mathbf{\Phi}_{\mathrm{t},1},\ldots,\mathbf{\Phi}_{\mathrm{t},G}),
\end{aligned}
\label{eq:group_blkdiag}
\end{equation}
where $\mathbf{\Phi}_{\mathrm{t/r},g}\in\mathbb{C}^{\bar{M}\times\bar{M}},\forall g\in\mathcal{G}$
satisfies the constraint 
\begin{equation}
\mathbf{\Phi}_{\mathrm{r},g}^{H}\mathbf{\Phi}_{\mathrm{r},g}+\mathbf{\Phi}_{\mathrm{t},g}^{H}\mathbf{\Phi}_{\mathrm{t},g}=\mathbf{I}_{\bar{M}},\forall g\in\mathcal{G}.\label{eq:Group_connected}
\end{equation}

\subsection{Mode Analysis}

According to the proportion of energy split for reflection and transmission, the BD-RIS is able to realize the following three different modes \cite{JXuSTAR,SZhang}:

\subsubsection{Reflective Mode}

The BD-RIS only reflects signals towards the same side as the transmitter, i.e., $\mathbf{\Phi}_{\mathrm{t}}=\mathbf{0}$, $\mathbf{\Phi}_{\mathrm{r}}^{H}\mathbf{\Phi}_{\mathrm{r}}=\mathbf{I}_{M}$.

\subsubsection{Transmissive Mode}

Incident signals only penetrate the BD-RIS, i.e., $\mathbf{\Phi}_{\mathrm{r}}=\mathbf{0}$,
$\mathbf{\Phi}_{\mathrm{t}}^{H}\mathbf{\Phi}_{\mathrm{t}}=\mathbf{I}_{M}$.

\subsubsection{Hybrid Reflective and Transmissive Mode}
Incident signals can both reflect from and transmit through the BD-RIS, which leads to a dual function of reflection and transmission\footnote{Note that reflective/transmissive mode only is the special case of the hybrid transmissive and reflective mode, and the hybrid mode can efficiently utilize the resources.}, i.e.,
$\mathbf{\Phi}_{\mathrm{r}}\ne\mathbf{0}$, $\mathbf{\Phi}_{\mathrm{t}}\ne\mathbf{0}$,
$\mathbf{\Phi}_{\mathrm{r}}^{H}\mathbf{\Phi}_{\mathrm{r}}+\mathbf{\Phi}_{\mathrm{t}}^{H}\mathbf{\Phi}_{\mathrm{t}}=\mathbf{I}_{M}$.

\subsection{Unified Architecture and Mode}

Combining three different BD-RIS architectures and three modes, there are in total nine cases. For clarity, the BD-RIS models for the nine cases are summarized in Table \ref{Table_1}.

\begin{remark} 
    In \cite{SShen}, an $M$-element RIS is modeled as $M$ antennas connected to an $M$-port reconfigurable impedance network. In this work, we further generalize the RIS model by modeling an $M$-cell RIS as 2$M$ antennas connected to a $2M$-port network.
    Namely, each RIS ``element'' in \cite{SShen} contains a single antenna, while each RIS ``cell'' in this work contains two back to back placed uni-directional antennas, which can support reflective and transmissive modes.
    Particularly, when the RIS works in the reflective mode, our proposed model boils down to the model in \cite{SShen}, as shown in the first row (starting from ``Reflective'') of Table \ref{Table_1}.
    Prior works on conventional RIS \cite{YHan,RZhang,YLiu2021,BDi,QWu,JXu,ZLi,BZhengMulti-IRS,WMei,Kris,YangZhao,BZheng2020,CLiu} is a subset of group/fully-connected RIS \cite{SShen} and also a special case of our model as summarized in Table \ref{Table_1}. 
\end{remark}

\begin{remark} 
    There are twofold differences between STAR-RIS/IOS \cite{JXuSTAR,SZhang,HZhang,JXuHybridRIS} and our proposed model. On one hand, STAR-RIS/IOS, which is essentially the CW-SC hybrid BD-RIS as shown in Table \ref{Table_1}, is a particular instance of the proposed model when there are no inter-cell connections. On the other hand, we go beyond the architecture of STAR-RIS/IOS and propose more general and flexible CW-GC/FC architectures. The proposed CW-GC/WC architectures are built on the multi-port reconfigurable impedance network \cite{SShen}, which cannot be derived based on the modeling of STAR-RIS/IOS \cite{JXuSTAR,SZhang,HZhang,JXuHybridRIS} due to the lack of a rigorous analysis using network theory.
\end{remark}

\begin{figure}
    \centering
    \includegraphics[height=1.7 in]{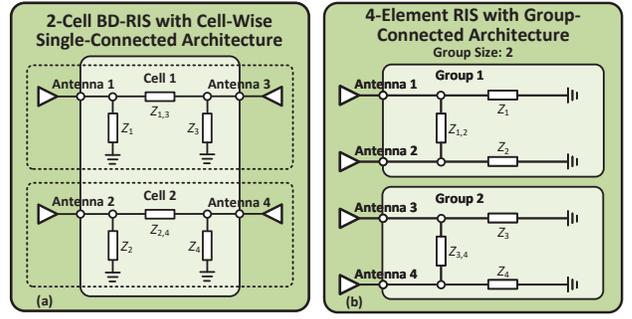}
    \caption{(a) 2-cell CW-SC BD-RIS and (b) 4-element group-connected RIS.}\label{fig:gc_relationship}
\end{figure}

\begin{remark}
    Given that the RIS can be generally modeled as multiple antennas connected to a multi-port reconfigurable impedance network, the circuit topology of the reconfigurable impedance network in STAR-RIS is essentially the same as that in the group-connected RIS with group size 2 \cite{SShen}. To facilitate understanding, we provide a simple example in Fig. \ref{fig:gc_relationship} to show the equivalence in terms of circuit topologies between the CW-SC 2-cell BD-RIS and group-connected 4-elements BD-RIS with group size 2. 
    The difference is that all the antennas of the group-connected BD-RIS with group size 2 are placed in the same direction, while antennas of the CW-SC BD-RIS are divided into two sectors and back to back placed. 
    Therefore, intra-cell connection together with the antenna arrangements supports ``modes'' while inter-cell connection refers to the ``architectures'' among different cells. 
\end{remark}

In the following section, we will focus on the beamforming design for an BD-RIS-aided wireless communication system so as to investigate and compare the nine cases of BD-RIS.

\vspace{0.6 cm}

\section{Joint Transmit Precoder and BD-RIS Matrix Design}

\subsection{Problem Formulation}

We consider an RIS-aided MU-MISO system with an $N$-antenna base station (BS), an $M$-cell BD-RIS, and $K$ single-antenna users (including $K_\mathrm{r}$ reflective users and $K_\mathrm{t}$ transmissive ones) based on Section II.
In this section, we have the following assumptions: \textit{i)} We assume direct links between the BS and users are blocked so that there are only BS-RIS-user links\footnote{Herein we ignore direct links for simplicity. However, our proposed algorithms still work if the direct link exists.}. \textit{ii)} We assume exact and instantaneous channel state information (CSI) is available at the BS, and our proposed design provides an upper bound on the performance of practical systems.

Let $\mathbf{s} \triangleq [s_1, \ldots, s_K]^T \in \mathbb{C}^{K}$ be the transmit symbol vector, $\mathbb{E}\{\mathbf{s}\mathbf{s}^H\} = \mathbf{I}_{K}$. Transmit symbols are first precoded at the BS by a precoder matrix $\mathbf{W} \triangleq [\mathbf{w}_1, \ldots, \mathbf{w}_{K}] \in \mathbb{C}^{N\times K}$, where $\mathbf{w}_{k} \in \mathbb{C}^N$ is the precoding vector for user $k$, $\forall k \in \mathcal{K}$. Then they are up-converted to the radio frequency (RF) domain via $N$ RF chains. After propagating through the RIS-aided channels, signals are corrupted by additive Gaussian white noise (AGWN). Thus, the received signal for each user is
\begin{equation}
    \begin{aligned}
        y_{k} =&\mathbf{h}_{k}^H\mathbf{\Phi}_i\mathbf{G}\mathbf{W}\mathbf{s} + n_{k},\\
        =&\mathbf{h}_{k}^H\mathbf{\Phi}_i \mathbf{G}\mathbf{w}_{k}s_{k} + \mathbf{h}_{k}^H\mathbf{\Phi}_i\mathbf{G}\sum_{p \in \mathcal{K}, p \ne k}\mathbf{w}_{p}s_{p}\\
        &+ n_{k}, \forall i \in \{\mathrm{t,r}\}, \forall k \in \mathcal{K}_i,
    \end{aligned}
\end{equation}
where $\mathbf{h}_{k} \in \mathbb{C}^{M}$ denotes the channel vector between the RIS and user $k$, and $n_{k} \sim \mathcal{CN}(0, \sigma_{k}^2)$ denotes the AGWN, $\forall k \in \mathcal{K}$.

Define $\tilde{\mathbf{h}}_{k} \triangleq (\mathbf{h}_{k}^H\mathbf{\Phi}_i\mathbf{G})^H$, $\forall i \in \{\mathrm{t,r}\}$, $\forall k \in \mathcal{K}_i$. Then the signal-to-interference-plus-noise ratio (SINR) for each user can be calculated as
\begin{equation}
    \begin{aligned}
        \gamma_{k} = &\frac{|\tilde{\mathbf{h}}_{k}^H \mathbf{w}_{k}|^2}{\sum_{p \in \mathcal{K}, p\ne k}|\tilde{\mathbf{h}}_{k}^H \mathbf{w}_{p}|^2 + \sigma_{k}^2}, \forall i \in \{\mathrm{t,r}\}, \forall k \in \mathcal{K}_i.
    \end{aligned}
\end{equation}

Our goal is to jointly design the transmit precoder and BD-RIS matrix to maximize the sum-rate for the MU-MISO system, subject to power constraint and the BD-RIS constraint shown in Table \ref{Table_1}. Therefore, the problem can be formulated as
\begin{subequations}
    \label{eq:problem0}
    \begin{align}
        \label{eq:obj0}
        \max_{\mathbf{W}, \mathbf{\Phi}_\mathrm{t}, \mathbf{\Phi}_\mathrm{r}} ~~ &f_o(\mathbf{W}, \mathbf{\Phi}_\mathrm{t}, \mathbf{\Phi}_\mathrm{r}) = \sum_{k\in \mathcal{K}}\log_2(1 + \gamma_{k})\\
        \label{eq:p0_b}
        \mathrm{s.t.} ~~~~ & \mathbf{\Phi}_\mathrm{t} ~ \mathrm{and}~ \mathbf{\Phi}_\mathrm{r}~ \mathrm{satisfy ~ Table ~\ref{Table_1}},\\
        \label{eq:p0_c}
        &\|\mathbf{W}\|_F^2 \le P,
    \end{align}
\end{subequations}
where $P$ is the total transmit power at the BS.
Problem (\ref{eq:problem0}) is difficult to solve due to the complex form of the objective and the non-convex constraints of the BD-RIS.
To tackle this difficulty, in the following subsections, we attempt to first transform problem (\ref{eq:problem0}) into a more tractable multi-variable/block optimization based on fractional programming theory and then iteratively cope with each block.

\subsection{Overview of the Joint Design Framework}

We start by taking the fractional terms $\gamma_{k}, \forall k \in \mathcal{K}$ out of the $\log(\cdot)$ function in the original objective $f_o(\mathbf{W},\mathbf{\Phi}_\mathrm{t},\mathbf{\Phi}_\mathrm{r})$. Based on the \textit{Lagrangian Dual Transform} \cite{FPI}, \cite{FPII},  $f_o(\mathbf{W},\mathbf{\Phi}_\mathrm{t},\mathbf{\Phi}_\mathrm{r})$ can be equivalently transformed into the following form:
\begin{equation}
    \begin{aligned}
        f_\iota(\mathbf{W},\mathbf{\Phi}_\mathrm{t},\mathbf{\Phi}_\mathrm{r}, {\bm \iota}) =& \sum_{k\in\mathcal{K}} \Bigg(\log_2(1 + \iota_{k}) - \iota_{k}\\
        &+ \frac{(1 + \iota_{k})|\tilde{\mathbf{h}}_{k}^H\mathbf{w}_{k}|^2}{\sum_{p\in\mathcal{K}} |\tilde{\mathbf{h}}_{k}^H\mathbf{w}_{p}|^2 + \sigma_{k}^2}\Bigg),
        \label{eq:f_l}
    \end{aligned}
\end{equation}
where ${\bm \iota} \triangleq [\iota_{1}, \ldots, \iota_{K}]^T \in \mathbb{R}^K$ is an auxiliary vector. It can be seen from (\ref{eq:f_l}) that we remove the original fractional terms in the $\log(\cdot)$ function of $f_o(\mathbf{W},\mathbf{\Phi}_\mathrm{t},\mathbf{\Phi}_\mathrm{r})$, but introduce a summation of new fractional terms. The main difference is that these new fractional terms are independent of the $\log(\cdot)$ function and thus more tractable. Then we can apply \textit{Quadratic Transform} \cite{FPI}, \cite{FPII} to transform these fractional parts into integral expressions and reformulate the objective function $f_\iota(\mathbf{W},\mathbf{\Phi}_\mathrm{t},\mathbf{\Phi}_\mathrm{r}, {\bm \iota})$ as:
\begin{equation}
    \begin{aligned}
        &f_\tau(\mathbf{W},\mathbf{\Phi}_\mathrm{t},\mathbf{\Phi}_\mathrm{r}, {\bm \iota}, {\bm \tau}) = \sum_{k\in\mathcal{K}} \Big(\log_2(1 + \iota_{k}) - \iota_{k} \\
        &+ 
        2\sqrt{1 + \iota_{k}}\Re\{\tau_{k}^*\tilde{\mathbf{h}}_{k}^H\mathbf{w}_{k}\} 
       - |\tau_{k}|^2 \sum_{p\in\mathcal{K}} \left(|\tilde{\mathbf{h}}_{k}^H\mathbf{w}_{p}|^2 + \sigma_{k}^2\right)\Big),
        \label{eq:f_t}
    \end{aligned}
\end{equation}
where ${\bm \tau} \triangleq [\tau_{1}, \ldots, \tau_{K}]^T \in \mathbb{C}^K$ denotes another auxiliary vector. Now with newly introduced two auxiliary vectors ${\bm \iota}$ and ${\bm \tau}$, the original problem (\ref{eq:problem0}) can be transformed into:
\begin{subequations}
    \label{eq:problem1}
    \begin{align}
        \max_{\mathbf{W}, \mathbf{\Phi}_\mathrm{t}, \mathbf{\Phi}_\mathrm{r},{\bm \iota}, {\bm \tau}} ~~ &f_\tau(\mathbf{W}, \mathbf{\Phi}_\mathrm{t}, \mathbf{\Phi}_\mathrm{r},{\bm \iota}, {\bm \tau})\\
        \label{eq:p1_b}
        \mathrm{s.t.} ~~~~~~ & \textrm{(\ref{eq:p0_b}), (\ref{eq:p0_c})}.
    \end{align}
\end{subequations}

\begin{algorithm}[t]
    \caption{Joint Transmit Precoder and BD-RIS Design}
    \label{alg:fp}
    \begin{algorithmic}[1]
        \REQUIRE $\mathbf{h}_{i,k}$, $\forall i \in \{\mathrm{t,r}\}$, $\forall k \in \mathcal{K}_i$, $\mathbf{G}$, $P$.
        \ENSURE $\mathbf{\Phi}_\mathrm{t}^\star$, $\mathbf{\Phi}_\mathrm{r}^\star$, $\mathbf{W}^\star$.
            \STATE {Initialize $\mathbf{\Phi}_\mathrm{t}$, $\mathbf{\Phi}_\mathrm{r}$, $\mathbf{W}$, $\mathcal{D}_1, \ldots, \mathcal{D}_G$}.
            \WHILE {no convergence of objective (\ref{eq:obj0})}
                \STATE {Update ${\bm \iota}^\star$ by (\ref{eq:iota}) in Section IV-C.}
                \STATE {Update ${\bm \tau}^\star$ by (\ref{eq:tau}) in Section IV-C.}
                \STATE {Update $\mathbf{W}^\star$ by (\ref{eq:w}) in Section IV-D.}
                \STATE {Update $\mathbf{\Phi}_\mathrm{t/r}^\star$ by solving problem (\ref{eq:sub_phi}) in Section IV-E.}
            \ENDWHILE
            \STATE {Return $\mathbf{\Phi}_\mathrm{t/r}^\star$, $\mathbf{W}^\star$.}
    \end{algorithmic}
\end{algorithm}

Problem (\ref{eq:problem1}) is a typical multi-variable/block problem. A well-known direction is to solve it based on block coordinate descent (BCD) iterative algorithms \cite{DBertsekas}. Given appropriate initial values of $\mathbf{W}$ and $\mathbf{\Phi}_\mathrm{t/r}$, we iteratively update the above blocks until convergence.
The proposed joint design framework is summarized in Algorithm \ref{alg:fp}.

In the following subsections, we will decompose problem (\ref{eq:problem1}) into several sub-problems in an iterative manner and discuss the solution for each block in detail. Specifically, solutions to two auxiliary vectors, i.e., blocks ${\bm \iota}$ and ${\bm \tau}$, will be presented in Section IV-C. Then, the solution to the transmit precoder, i.e., block $\mathbf{W}$, will be provided in Section IV-D. Finally, the solution to the BD-RIS beamformer, i.e., block $\{\mathbf{\Phi}_\mathrm{t}, \mathbf{\Phi}_\mathrm{r}\}$, will be discussed in Section IV-E.

\subsection{Auxiliary Vectors: Blocks ${\bm \iota}$ and ${\bm \tau}$} 
When $\mathbf{W}$, $\mathbf{\Phi}_\mathrm{t}$, $\mathbf{\Phi}_\mathrm{r}$, and ${\bm \tau}$ (or ${\bm \iota}$) are fixed, the sub-problem with respect to ${\bm \iota}$ (or ${\bm \tau}$) is an unconstrained convex optimization, whose solution can be easily derived by setting $\frac{\partial f_\tau(\mathbf{W}, \mathbf{\Phi}_\mathrm{t}, \mathbf{\Phi}_\mathrm{r},{\bm \iota}, {\bm \tau})}{\partial {\bm \iota}} = \mathbf{0}$ (or $\frac{\partial f_\tau(\mathbf{W}, \mathbf{\Phi}_\mathrm{t}, \mathbf{\Phi}_\mathrm{r},{\bm \iota}, {\bm \tau})}{\partial {\bm \tau}} = \mathbf{0}$). Then we can obtain the following optimal solutions for each auxiliary variables $\iota_{k}$ and $\tau_{k}$ as:
\begin{equation}
    \label{eq:iota}
    \iota_{k}^\star = \gamma_{k}, \forall k \in \mathcal{K},
\end{equation}
\begin{equation}
    \label{eq:tau}
        \tau_{k}^\star = \frac{\sqrt{1 + \iota_{k}}\tilde{\mathbf{h}}_{k}^H\mathbf{w}_{k}}{\sum_{p \in \mathcal{K}}|\tilde{\mathbf{h}}_{k}^H\mathbf{w}_{p}|^2 + \sigma_{k}^2}, \forall k \in \mathcal{K}.
\end{equation}

\subsection{Transmit Precoder: Block $\mathbf{W}$} 
With given $\mathbf{\Phi}_\mathrm{t}$, $\mathbf{\Phi}_\mathrm{r}$, ${\bm \tau}$, and ${\bm \iota}$, the sub-problem with respect to $\mathbf{W}$ is given by:
\begin{subequations}
    \label{eq:sub_w}
    \begin{align}
        \max_{\mathbf{W}} ~ &\sum_{k\in\mathcal{K}}\Big(
        2\sqrt{1 + \iota_{k}}\Re\{\bar{\mathbf{h}}_{k}^H\mathbf{w}_{k}\} - \mathbf{w}_{k}^H\Big(\sum_{p\in\mathcal{K}} \bar{\mathbf{h}}_{p}\bar{\mathbf{h}}_{p}^H\Big)\mathbf{w}_{k}\Big)\\
        \label{eq:sub_w_constraint}
        \mathrm{s.t.} ~~ &\|\mathbf{W}\|_F^2 \le P,
    \end{align}
\end{subequations}
where $\bar{\mathbf{h}}_{k} \triangleq \tau_{k}\tilde{\mathbf{h}}_{k}, \forall k \in \mathcal{K}$.
Since both the objective function and the constraint of problem (\ref{eq:sub_w}) are convex, we can use some classical optimization methods to find its optimal solution. Here we adopt the Lagrange multiplier method and introduce a multiplier $\lambda \ge 0$ for the power constraint (\ref{eq:sub_w_constraint}).
Then by checking the first-order optimum condition, we can obtain the optimal solution of precoder $\mathbf{W}$:
\begin{equation}
    \label{eq:w}
        \mathbf{w}_{k}^\star = \left(\sum_{p\in\mathcal{K}} \bar{\mathbf{h}}_{p}\bar{\mathbf{h}}_{p}^H + \lambda^\star\mathbf{I}_N\right)^{-1}\times\sqrt{1 + \iota_{k}}\bar{\mathbf{h}}_{k}, \forall k \in \mathcal{K},
\end{equation}
where $\lambda^\star$ can be obtained by a simple bisection search.

\subsection{BD-RIS Matrix: Block $\{\mathbf{\Phi}_\mathrm{t}, \mathbf{\Phi}_\mathrm{r}\}$} 
When $\mathbf{W}$, ${\bm \tau}$, and ${\bm \iota}$ are determined, the sub-problem with respect to $\mathbf{\Phi}_\mathrm{t}$ and $\mathbf{\Phi}_\mathrm{r}$ is written as
\begin{subequations}\label{eq:sub_phi}
    \begin{align}
        \max_{\mathbf{\Phi}_\mathrm{t},\mathbf{\Phi}_\mathrm{r}} ~ &\sum_{i\in\{\mathrm{t,r}\}}\sum_{k\in\mathcal{K}_i}\Big(
        2\Re\{\tilde{\tau}_{k}^*\mathbf{h}_{k}^H\mathbf{\Phi}_i\mathbf{g}_{k}\} - |\tau_{k}|^2 \sum_{p\in\mathcal{K}} |\mathbf{h}_{k}^H\mathbf{\Phi}_i\mathbf{g}_{p}|^2\Big)\\
        \mathrm{s.t.} ~~ &\mathbf{\Phi}_\mathrm{t} ~ \mathrm{and}~ \mathbf{\Phi}_\mathrm{r}~ \mathrm{satisfy ~ Table ~\ref{Table_1}},
    \end{align}
\end{subequations}
where $\tilde{\tau}_k = \sqrt{1 + \iota_{k}}\tau_{k}$, $\mathbf{g}_{k} \triangleq \mathbf{G}\mathbf{w}_{k}, \forall k \in \mathcal{K}$. 
It can be easily observed from Table I that both CW-SC and CW-FC architectures are special cases of the CW-GC architecture.
Therefore, in the following section, we will first provide a general solution for the CW-GC case, which can easily boil down to other cases, and then propose an efficient algorithm for the CW-SC case to further reduce the complexity.

\subsubsection{A General Solution for CW-GC BD-RIS} 
Problem (\ref{eq:sub_phi}) can be written as 
\begin{subequations}\label{eq:sub_phi_group}
    \begin{align}
    \label{eq:obj_phi_group}
        \max_{\mathbf{\Phi}_\mathrm{t},\mathbf{\Phi}_\mathrm{r}} &\sum_{i\in\{\mathrm{t,r}\}}\Big(
        2\Re\{\mathsf{Tr}(\mathbf{\Phi}_i\mathbf{X}_i)\} - \mathsf{Tr}(\mathbf{\Phi}_i\mathbf{Y}\mathbf{\Phi}_i^H\mathbf{Z}_i)\Big)\\
        \mathrm{s.t.} ~ &\mathbf{\Phi}_{\mathrm{t},g}^H\mathbf{\Phi}_{\mathrm{t},g} + \mathbf{\Phi}_{\mathrm{r},g}^H\mathbf{\Phi}_{\mathrm{r},g} = \mathbf{I}_{\bar{M}}, \forall g \in \mathcal{G},\\
        &\mathbf{\Phi}_i = \mathsf{blkdiag}(\mathbf{\Phi}_{i,1}, \ldots, \mathbf{\Phi}_{i,G}), \forall i \in \{\mathrm{t,r}\},
    \end{align}
\end{subequations}
where $\mathbf{X}_i \in \mathbb{C}^{M\times M}$, $\mathbf{Y}\in \mathbb{C}^{M\times M}$, and $\mathbf{Z}_i \in \mathbb{C}^{M\times M}$ are respectively defined as
\begin{equation}
\label{eq:XYZ}
    \begin{aligned}
        &\mathbf{X}_i \triangleq \sum_{k \in \mathcal{K}_i}\tilde{\tau}_{k}^*\mathbf{g}_{k}\mathbf{h}_{k}^H,~~\mathbf{Y} \triangleq \sum_{p\in\mathcal{K}} \mathbf{g}_{p}\mathbf{g}_{p}^H,\\
        &\mathbf{Z}_i \triangleq \sum_{k \in \mathcal{K}_i} |\tau_{k}|^2\mathbf{h}_{k}\mathbf{h}_{k}^H, \forall i \in \{\mathrm{t,r}\}.
    \end{aligned}
\end{equation}

It is worth noting that there are quadratic terms in problem (\ref{eq:sub_phi_group}), which makes different groups of $\mathbf{\Phi}_{\mathrm{t/r},g}$ mingle with each other. 
To efficiently solve this problem, we attempt to split out one pair $\mathbf{\Phi}_{\mathrm{t},g}$ and $\mathbf{\Phi}_{\mathrm{r},g}$ for group $g$ from the objective function with fixed other pairs, and focus on the design of each pair. To this end, we first split objective (\ref{eq:obj_phi_group}) group-by-group as:
\begin{equation}
\begin{aligned}
    &\sum_{i\in\{\mathrm{t,r}\}}\Big(
        2\Re\{\mathsf{Tr}(\mathbf{\Phi}_i\mathbf{X}_i)\} - \mathsf{Tr}(\mathbf{\Phi}_i\mathbf{Y}\mathbf{\Phi}_i^H\mathbf{Z}_i)\Big)\\
    = &\sum_{i\in\{\mathrm{t,r}\}}\Bigg(2\sum_{q=1}^G\Re\{\mathsf{Tr}(\mathbf{\Phi}_{i,q} \mathbf{X}_{i,q})\}\\
     &- \mathsf{Tr}\left(\sum_{p = 1}^G\mathbf{\Phi}_{i,p}\sum_{q = 1}^G \mathbf{Y}_{p,q}\mathbf{\Phi}_{i,q}^H\mathbf{Z}_{i,q,p} \right)\Bigg),
\end{aligned}
\end{equation}
where $\mathbf{X}_{i,q} = [\mathbf{X}_i]_{(q-1)\bar{M}+1:q\bar{M},(q-1)\bar{M}+1:q\bar{M}} \in \mathbb{C}^{\bar{M}\times \bar{M}}$, $\mathbf{Y}_{p,q} = [\mathbf{Y}]_{(p-1)\bar{M}+1:p\bar{M},(q-1)\bar{M}+1:q\bar{M}}\in \mathbb{C}^{\bar{M}\times \bar{M}}$, and $\mathbf{Z}_{i,q,p} = [\mathbf{Z}_i]_{(q-1)\bar{M}+1:q\bar{M},(p-1)\bar{M}+1:p\bar{M}}\in \mathbb{C}^{\bar{M}\times \bar{M}}$, $\forall p, q \in \mathcal{G}$. Then, when we focus on one pair $\mathbf{\Phi}_{\mathrm{t},g}$ and $\mathbf{\Phi}_{\mathrm{r},g}$ with fixed other pairs, the corresponding sub-objective function has the following form:
\begin{equation}
    \begin{aligned}
        \label{eq:obj_phi_group_m}
        &f_g(\mathbf{\Phi}_{\mathrm{t},g}, \mathbf{\Phi}_{\mathrm{r},g}) = \sum_{i\in\{\mathrm{t,r}\}}\Bigg( \mathsf{Tr}(\mathbf{\Phi}_{i,g} \mathbf{Y}_{g,g}\mathbf{\Phi}_{i,g}^H\mathbf{Z}_{i,g,g})\\
        &- 2\Re\Big\{\mathsf{Tr}\Big(\mathbf{\Phi}_{i,g}\big(\underbrace{\mathbf{X}_{i,g} - \sum_{p \ne g}\mathbf{Y}_{g,p}\mathbf{\Phi}_{i,p}^H\mathbf{Z}_{i,p,g}}_{= \tilde{\mathbf{X}}_{i,g}}\big)\Big)\Big\}\Bigg).
    \end{aligned}
\end{equation}
Define matrices $\mathbf{\Phi}_{g} \triangleq [\mathbf{\Phi}_{\mathrm{t},g}^H, \mathbf{\Phi}_{\mathrm{r},g}^H]^H \in \mathbb{C}^{2\bar{M}\times \bar{M}}$, $\tilde{\mathbf{X}}_{g} \triangleq [\tilde{\mathbf{X}}_{\mathrm{t},g},\tilde{\mathbf{X}}_{\mathrm{r},g}] \in \mathbb{C}^{\bar{M}\times 2\bar{M}}$, and $\mathbf{Z}_{g} \triangleq \mathsf{blkdiag} (\mathbf{Z}_{\mathrm{t},g,g}, \mathbf{Z}_{\mathrm{r},g,g}) \in \mathbb{C}^{2\bar{M}\times 2\bar{M}}$. Then the sub-problem with respect to $\mathbf{\Phi}_{g}$ can be formulated as follows:
\begin{subequations}
    \label{eq:sub_phi_group_g2}
    \begin{align}
        \min_{\mathbf{\Phi}_{g}} ~ &\tilde{f}_g(\mathbf{\Phi}_{g}) = \mathsf{Tr}(\mathbf{\Phi}_{g} \mathbf{Y}_{g,g}\mathbf{\Phi}_{g}^H\mathbf{Z}_{g}
        - 2\Re\{\mathbf{\Phi}_{g}\tilde{\mathbf{X}}_{g}\})\\
        \label{eq:group_unitary_constraint}
        \mathrm{s.t.} ~ & \mathbf{\Phi}_{g}^H\mathbf{\Phi}_{g} = \mathbf{I}_{\bar{M}}, \forall g \in \mathcal{G}.
    \end{align}
\end{subequations}

The main challenge of solving problem (\ref{eq:sub_phi_group_g2}) is the unitary constraint (\ref{eq:group_unitary_constraint}). One possible solution is to do some relaxations and solve the relaxed problem. However, it is inevitable that there will be performance loss since the result is not based on the original problem. Moreover, when the value of $\mathbf{\Phi}_{g}$ is not that ``good'', the convergence of the overall BCD procedure might not be guaranteed. In order to get a ``good'' solution of problem (\ref{eq:sub_phi_group_g2}), we adopt a manifold algorithm \cite{PAAbsil}, whose main idea is to construct all available solutions of the problem as a manifold, and then transform the original constrained problem on the Euclidean space into an unconstrained one on the manifold space. Then many algorithms on the Euclidean space, e.g., Trust-Region (TR), Conjugate-Gradient (CG), and Broyden-Fletcher-Goldfarb-Shanno (BFGS) methods \cite{NBoumal}, can be easily extended to the manifold space with some necessary projections. In the following we will introduce the procedure of the manifold algorithm by applying the CG method.

\textit{Manifold Construction:} Constraint (\ref{eq:group_unitary_constraint}) forms a $2\bar{M}^2$-dimensional complex Stiefel manifold \cite{NBoumal}, i.e.,
\begin{equation}
    \label{eq:manifold}
    \mathcal{M}_g = \{\mathbf{\Phi}_{g} \in \mathbb{C}^{2\bar{M}\times \bar{M}}: \mathbf{\Phi}_{g}^H\mathbf{\Phi}_{g} = \mathbf{I}_{\bar{M}}\}, \forall g \in \mathcal{G},
\end{equation}
and then problem (\ref{eq:sub_phi_group_g2}) becomes an unconstrained optimization on the Stiefel manifold, i.e.,
\begin{equation}
    \mathbf{\Phi}_{g}^\star = \arg\min_{\mathbf{\Phi}_{g} \in \mathcal{M}_g} \tilde{f}_g(\mathbf{\Phi}_{g}), \forall g \in \mathcal{G}.
\end{equation}
Here we should explain that a manifold is a topological space which ``locally'' resembles the Euclidean space. To be specific, the moving direction of a point on the manifold is referred to as a tangent vector. All these tangent vectors at this point, which include all possible directions this point can move to, form the tangent space. Each tangent space can be regarded as a Euclidean space which has one tangent vector, namely, Riemannian gradient, pointing to the direction that the objective function decreases fastest \cite{PAAbsil}. The tangent space of the manifold (\ref{eq:manifold}) at point $\mathbf{\Phi}_g$ is given as
\begin{equation}
    \label{eq:tangent_space}
    \mathsf{T}_{\mathbf{\Phi}_{g}}\mathcal{M}_g
    = \{\mathbf{T}_{g} \in \mathbb{C}^{2\bar{M}\times \bar{M}}: \Re\{\mathbf{\Phi}_{g}^H\mathbf{T}_{g}\} = \mathbf{0}_{\bar{M}}\}, \forall g \in \mathcal{G}.
\end{equation}

\textit{Riemannian Gradient:} In the CG method, the Euclidean gradient is required to calculate the Riemannian gradient. Therefore, we first calculate the Euclidean gradient of $\tilde{f}_g(\mathbf{\Phi}_{g})$
\begin{equation}
    \triangledown \tilde{f}_g(\mathbf{\Phi}_{g}) = 2\mathbf{Z}_{g}\mathbf{\Phi}_{g}\mathbf{Y}_{g} - 2\mathbf{X}_{g}^H, \forall g \in \mathcal{G}.
\end{equation}
Then the Riemannian gradient can be calculated by projecting the Euclidean gradient onto the tangent space \cite{PAAbsil}:
\begin{equation}
    \begin{aligned}
        \triangledown_{\mathcal{M}_g}\tilde{f}_g(\mathbf{\Phi}_{g})
        = &\mathsf{Pr}_{\mathbf{\Phi}_{g}}(\triangledown \tilde{f}_g(\mathbf{\Phi}_{g}))\\
        = &\triangledown \tilde{f}_g(\mathbf{\Phi}_{g}) - \mathbf{\Phi}_{g}\mathsf{chdiag}(\mathbf{\Phi}_{g}^H \triangledown \tilde{f}_g(\mathbf{\Phi}_{g})), \forall g \in \mathcal{G},
    \end{aligned}
\end{equation}
where $\mathsf{Pr}_{\mathbf{\Phi}_{g}}(\cdot)$ denotes the projection function, and $\mathsf{chdiag}(\cdot)$ chooses all diagonal elements of a matrix to construct a diagonal matrix. 
Now we can apply the CG method. At the $v$-th iteration of the CG method, we successively do the following steps: 

\begin{itemize}
    \item[S1:] Find the descent direction:
    \begin{equation}
        \label{eq:xi}
        \mathbf{\Xi}_g^v = -\triangledown_{\mathcal{M}_g}\tilde{f}_g(\mathbf{\Phi}_{g}^v) + \mu_g^v\mathsf{Pr}_{\mathbf{\Phi}_{g}^v}(\mathbf{\Xi}_g^{v-1}), \forall g \in \mathcal{G},
    \end{equation}
    where $\mu_g^v$ denotes the CG update parameter. There are many choices for updating this parameter, e.g., Fletcher-Reeves formula, Polak-Ribi$\grave{\textrm{e}}$re formula, and Hestenes-Stiefel formula \cite{WWHager}. Here we adopt a Riemannian version of the Polak-Ribi$\grave{\textrm{e}}$re formula, which is given by
    \begin{equation}
        \label{eq:mu}
        \begin{aligned}
            \mu_g^v = &
            \Re\Big\{\mathsf{Tr}\Big([\triangledown_{\mathcal{M}_g}\tilde{f}_g(\mathbf{\Phi}_{g}^v)]^H \big[\triangledown_{\mathcal{M}_g}\tilde{f}_g(\mathbf{\Phi}_{g}^v)\\
            &-\mathsf{Pr}_{\mathbf{\Phi}_{g}^v}(\triangledown_{\mathcal{M}_g}\tilde{f}_g(\mathbf{\Phi}_{g}^{v-1}))\big]\Big)\Big\}/\\
            &{\mathsf{Tr}\left([\triangledown_{\mathcal{M}_g}\tilde{f}_g(\mathbf{\Phi}_{g}^{v-1})]^H \triangledown_{\mathcal{M}_g}\tilde{f}_g(\mathbf{\Phi}_{g}^{v-1})\right)}, \forall g \in \mathcal{G}.
        \end{aligned}
    \end{equation}
    \item[S2:] Perform a retraction \cite{PAAbsil}:
    \begin{equation}
        \label{eq:retraction}
            \begin{aligned}
                &\mathbf{\Phi}_{g}^{v+1} = \mathsf{Retr}_{\mathbf{\Phi}_{g}^{v}}(\delta_g^v\mathbf{\Xi}_g^v)\\
                =& (\mathbf{\Phi}_{g}^{v} + \delta_g^v\mathbf{\Xi}_g^v)(\mathbf{I}_{\bar{M}} + (\delta_g^v)^2(\mathbf{\Xi}_g^v)^H\mathbf{\Xi}_g^v)^{-1/2}, \forall g \in \mathcal{G},
            \end{aligned}
        \end{equation}
    where $\delta_g^v$ is the step size and can be searched by backtracking algorithms \cite{PAAbsil}. $\mathsf{Retr}_{\mathbf{\Phi}_g^v}(\cdot)$ denotes the retraction function which maps a point in the tangent space into the manifold. 
\end{itemize}

Now with proper initial values, the (at least local) optimal $\mathbf{\Phi}_{g}^\star$, $\forall g$ can be obtained by iteratively updating $\delta_g^{v-1}$, $\mathbf{\Phi}_{g}^v$, $\mu_g^v$, and $\mathbf{\Xi}_g^v$ until convergence.
After solving problem (\ref{eq:sub_phi_group_g2}) for all groups, the optimal reflective and transmissive matrices for each group can be split from $\mathbf{\Phi}_{g}^\star$, i.e.,
\begin{equation}
    \mathbf{\Phi}_{\mathrm{t}, g}^\star = [\mathbf{\Phi}_{g}^\star]_{1:\bar{M},:}, ~~~ \mathbf{\Phi}_{\mathrm{r}, g}^\star = [\mathbf{\Phi}_{g}^\star]_{\bar{M} + 1 : 2\bar{M},:}, \forall g \in \mathcal{G}.
    \label{eq:group_opt_phy}
\end{equation}
The procedure of the above BD-RIS matrix design is summarized in Algorithm \ref{alg:MCG}.

\begin{algorithm}[!t]
    \caption{General Solution for CW-GC BD-RIS}
    \label{alg:MCG}
    \begin{algorithmic}[1]
        \REQUIRE $\mathbf{h}_{i,k}$, $\forall i \in \{\mathrm{t,r}\}$, $\forall k \in \mathcal{K}_i$, $\mathbf{G}$, ${\bm \iota}$, ${\bm \tau}$, $\mathbf{W}$, $\mathbf{\Phi}_\mathrm{t}$, $\mathbf{\Phi}_\mathrm{r}$, $G$.
        \ENSURE $\mathbf{\Phi}_\mathrm{t}^\star$, $\mathbf{\Phi}_\mathrm{r}^\star$.
        \STATE {Calculate $\mathbf{X}_\mathrm{t/r}$, $\mathbf{Y}$, $\mathbf{Z}_\mathrm{t/r}$ by (\ref{eq:XYZ})}.
        \STATE{Initialize $\mathbf{\Phi}^0 = [\mathbf{\Phi}_\mathrm{t}^H, \mathbf{\Phi}_\mathrm{r}^H]^H$, $\mathbf{\Xi}_g^0 = - \triangledown_{\mathcal{M}_g}\tilde{f}_g(\mathbf{\Phi}_{g}^0)$, $\forall g \in \mathcal{G}$.}
        \WHILE {no convergence of $\mathbf{\Phi}_\mathrm{t}$, $\mathbf{\Phi}_\mathrm{r}$}
            \FOR{$g = 1 : G$}
                \STATE{Set $v = 0$.}
                \WHILE {no convergence of $\|\triangledown_{\mathcal{M}_g}\tilde{f}_g(\mathbf{\Phi}_{g}^v)\|_F$ }
                    \STATE {$v = v + 1$.}
                    \STATE {Calculate $\delta_g^{v-1}$ by backtracking algorithms \cite{PAAbsil}.}
                    \STATE {Update $\mathbf{\Phi}_{g}^v$ by (\ref{eq:retraction}).}
                    \STATE {Update $\mu_g^v$ by (\ref{eq:mu}).}
                    \STATE {Update $\mathbf{\Xi}_g^v$ by (\ref{eq:xi}).}
                \ENDWHILE
                \STATE {Obtain $\mathbf{\Phi}_{g}^\star = \mathbf{\Phi}_{g}^{v+1}$.}
                \STATE {Obtain $\mathbf{\Phi}_{\mathrm{t/r},g}^\star$ by (\ref{eq:group_opt_phy}).}
            \ENDFOR  
        \ENDWHILE      
        \STATE {Return $\mathbf{\Phi}_\mathrm{t/r}^\star$ by (\ref{eq:group_blkdiag}).}
    \end{algorithmic}
\end{algorithm}

\begin{remark}
    The proposed general solution is also suitable for CW-FC and CW-SC cases:
    \begin{itemize}
        \item[C1:] For the CW-FC case ($G=1$), problem (\ref{eq:sub_phi}) can be rewritten as the following form:
        \begin{subequations}\label{eq:sub_phi_full}
            \begin{align}
                \min_{\overline{\mathbf{\Phi}}} ~~ &\mathsf{Tr}(\overline{\mathbf{\Phi}}\mathbf{Y}\overline{\mathbf{\Phi}}^H\mathbf{Z}) - 2\Re\{\mathsf{Tr}(\overline{\mathbf{\Phi}}\mathbf{X})\}\\
                \label{eq:full_constraint}
                \mathrm{s.t.} ~~ & \overline{\mathbf{\Phi}}^H\overline{\mathbf{\Phi}} = \mathbf{I}_M,
            \end{align}
        \end{subequations}    
        where $\overline{\mathbf{\Phi}} \triangleq [\mathbf{\Phi}_\mathrm{t}^H, \mathbf{\Phi}_\mathrm{r}^H]^H \in \mathbb{C}^{2M\times M}$, $\mathbf{X} \triangleq [\mathbf{X}_\mathrm{t},\mathbf{X}_\mathrm{r}] \in \mathbb{C}^{M\times 2M}$, and $\mathbf{Z} \triangleq \mathsf{blkdiag} (\mathbf{Z}_\mathrm{t}, \mathbf{Z}_\mathrm{r}) \in \mathbb{C}^{2M \times 2M}$. This problem can be solved by steps 6-15 in Algorithm \ref{alg:MCG}. 
        \item[C2:] CW-SC RIS matrix can be directly determined by Algorithm \ref{alg:MCG} with $G = M$. However, CW-SC RIS has a friendly characteristic, i.e., $\mathbf{\Phi}_\mathrm{t/r}$ are diagonal matrices, which can facilitate the design. Therefore, we propose a more efficient algorithm for the CW-SC case, which achieves a similar performance but has lower computational complexity compared to Algorithm \ref{alg:MCG}. 
    \end{itemize}
\end{remark}

\subsubsection{An Efficient Solution for CW-SC BD-RIS} In this case, each non-zero element of $\mathbf{\Phi}_\mathrm{t/r}$ can be modeled as $\phi_{\mathrm{t/r},m} = \sqrt{\alpha_{\mathrm{t/r},m}}e^{\jmath\theta_{\mathrm{t/r},m}}$, $\theta_{\mathrm{t/r},m} \in [0, 2\pi)$, $\alpha_{\mathrm{t},m} + \alpha_{\mathrm{r},m} = 1$, $\forall m \in \mathcal{M}$ \cite{JXuSTAR}. Define ${\bm \phi}_i \triangleq [\phi_{i,1}, \ldots, \phi_{i,M}]^T$, and $\mathbf{v}_{k,p} \triangleq (\mathbf{h}_{k}^H\mathsf{diag}(\mathbf{g}_{p}))^H$, $\forall p\in \mathcal{K}$, $\forall k \in \mathcal{K}_i$, $\forall i \in \{\mathrm{t,r}\}$. 
Problem (\ref{eq:sub_phi}) can be rewritten as
\begin{subequations}
    \label{eq:sub_phi_single}
    \begin{align}
        \label{eq:obj_phi_single}
        \max_{{\bm \phi}_\mathrm{t}, {\bm \phi}_\mathrm{r}} ~ &\sum_{i\in\{\mathrm{t,r}\}}\Big(
        2\Re\{\tilde{\mathbf{v}}_i^H{\bm \phi}_i\} - {\bm \phi}_i^H\mathbf{V}_i{\bm \phi}_i\Big),\\
        \mathrm{s.t.}~~& |\phi_{\mathrm{t},m}|^2 + |\phi_{\mathrm{r},m}|^2 = 1, \forall m \in \mathcal{M},
    \end{align}
\end{subequations}
with definitions
\begin{equation}
    \label{eq:v}
    \begin{aligned}
        \mathbf{V}_i &\triangleq \sum_{k \in \mathcal{K}_i}|\tau_{k}|^2 \sum_{p\in\mathcal{K}} \mathbf{v}_{k,p}\mathbf{v}_{k,p}^H, \\
        \tilde{\mathbf{v}}_i &\triangleq \sum_{k \in \mathcal{K}_i} \sqrt{1 + \iota_{k}}\mathbf{v}_{k,k}\tau_{k}, \forall i \in \{\mathrm{t,r}\}.
    \end{aligned}
\end{equation}
We split the objective and focus on the design for one pair, i.e., $\phi_{\mathrm{t},m}, \phi_{\mathrm{r},m}$, while fixing the others. Therefore, the sub-objective for $\phi_{\mathrm{t},m}$, and $\phi_{\mathrm{r},m}$ is given by
\begin{equation}
    \label{eq:sub_phi_single_m}
    \begin{aligned}
        z_m(&\phi_{\mathrm{t},m}, \phi_{\mathrm{r},m}) =\sum_{i\in\{\mathrm{t,r}\}}\Bigg([\mathbf{V}_i]_{m,m}|\phi_{i,m}|^2 \\
        &+ 2\Re\Big\{\underbrace{\Big(\sum_{n\ne m}[\mathbf{V}_i]_{m,n}\phi_{i,n} - [\tilde{\mathbf{v}}_i]_m\Big)}_{= \chi_{i,m}}\phi_{i,m}^*\Big\}\Bigg).
    \end{aligned}
\end{equation}
Recall $\phi_{i,m} = \sqrt{\alpha_{i,m}}e^{\jmath\theta_{i,m}}$, $\forall i \in \{\mathrm{t,r}\}$, $\forall m \in \mathcal{M}$, the sub-problem for $\phi_{\mathrm{t},m}$, and $\phi_{\mathrm{r},m}$ is
\begin{subequations}
    \label{eq:sub_phi_single_m2}
    \begin{align}
        \label{eq:z}
        \min_{\substack{\alpha_{\mathrm{t},m}, \alpha_{\mathrm{r},m}\\ \theta_{\mathrm{t},m}, \theta_{\mathrm{r},m}}} ~ &\sum_{i\in\{\mathrm{t,r}\}}\Big([\mathbf{V}_i]_{m,m}\alpha_{i,m}\\ 
        &+ 2|\chi_{i,m}|\sqrt{\alpha_{i,m}}\cos(\angle\chi_{i,m} - \theta_{i,m})\Big)\\
        \label{eq:amplitude_constraint}
        \mathrm{s.t.}~~~~&\alpha_{\mathrm{t},m} + \alpha_{\mathrm{r},m} = 1,\\
        & \theta_{\mathrm{t},m} \in [0,2\pi), \theta_{\mathrm{r},m} \in [0,2\pi).
    \end{align}
\end{subequations}

\begin{algorithm}[!t]
    \caption{Efficient Solution for CW-SC BD-RIS}
    \label{alg:Single}
    \begin{algorithmic}[1]
        \REQUIRE $\mathbf{h}_{i,k}$, $\forall i \in \{\mathrm{t,r}\}$, $\forall k \in \mathcal{K}_i$, $\mathbf{G}$, ${\bm \iota}$, ${\bm \tau}$, $\mathbf{W}$, $\mathbf{\Phi}_\mathrm{t}$, $\mathbf{\Phi}_\mathrm{r}$.
        \ENSURE $\mathbf{\Phi}_\mathrm{t}^\star$, $\mathbf{\Phi}_\mathrm{r}^\star$.
            \STATE {Calculate $\mathbf{V}_\mathrm{t/r}$, $\tilde{\mathbf{v}}_\mathrm{t/r}$ by (\ref{eq:v}).}
            \WHILE {no convergence of $\mathbf{\Phi}_\mathrm{t}$, $\mathbf{\Phi}_\mathrm{r}$ }
                \FOR {$m=1:M$}
                    \STATE {Update $\theta_{\mathrm{t/r}, m}^\star$ by (\ref{eq:theta}).}
                    \STATE {Update $\alpha_{\mathrm{t},m}^\star$ by golden-section search.}
                    \STATE {Calculate $\phi_{\mathrm{t/r},m}^\star$ by (\ref{eq:single_opt_phi}).}
                \ENDFOR
            \ENDWHILE
            \STATE {Return $\mathbf{\Phi}_\mathrm{t/r}^\star = \mathsf{diag}(\phi_{\mathrm{t/r}, M}^\star, \ldots, \phi_{\mathrm{t/r}, M}^\star)$.}
    \end{algorithmic}
\end{algorithm}

\textit{Phase Shift:} From problem (\ref{eq:sub_phi_single_m2}) we can observe that the phase shifts $\theta_{\mathrm{t/r},m}$ and amplitudes $\alpha_{\mathrm{t/r},m}$ can be separately determined. Therefore, we first directly let the term $\cos(\angle\chi_{i,m} - \theta_{i,m}) = -1$ to minimize the objective function. Then we can determine the optimal phase shifts:
\begin{equation}\label{eq:theta}
    \begin{aligned}
        \theta_{i,m}^\star = &\begin{cases}
        \angle\chi_{i,m} + \pi, &\angle\chi_{i,m} \in [0,\pi),\\
        \angle\chi_{i,m} - \pi, &\angle\chi_{i,m} \in [\pi,2\pi),
        \end{cases}\\
        &~~~~~~~~~~~\forall i \in \{\mathrm{t,r}\}, \forall m \in \mathcal{M}.
    \end{aligned}
\end{equation}

\textit{Amplitude:} With optimal phase shifts, objective (\ref{eq:z}) becomes a real-value optimization and problem (\ref{eq:sub_phi_single_m2}) can be formulated as a function of $\alpha_{\mathrm{t},m}$:
\begin{subequations}
    \label{eq:sub_amplitude_single_m2}
    \begin{align}
        \label{eq:obj_amplitude2}
        \min_{\alpha_{\mathrm{t},m}} ~ &\upsilon_m\alpha_{\mathrm{t},m}
        - 2|\chi_{\mathrm{t},m}|\sqrt{\alpha_{\mathrm{t},m}} - 2|\chi_{\mathrm{r},m}|\sqrt{1 - \alpha_{\mathrm{t},m}}\\
        \mathrm{s.t.}~~& \alpha_{\mathrm{t},m} \in (0,1),
    \end{align}
\end{subequations}
where $\upsilon_m = [\mathbf{V}_\mathrm{t}]_{m,m} - [\mathbf{V}_\mathrm{r}]_{m,m}$, $\forall m \in \mathcal{M}$. It can be proved that objective  (\ref{eq:obj_amplitude2}) is a convex function with $\alpha_{\mathrm{t},m} \in (0,1)$ (the proof of the convexity of objective (\ref{eq:obj_amplitude2}) is provided in the Appendix). Therefore, the minimum of problem (\ref{eq:sub_amplitude_single_m2}) is achieved at the minimum point. Although we cannot get a close-form solution due to the complicated form of objective (\ref{eq:obj_amplitude2}), we can apply efficient one-dimensional search methods, e.g., golden-section search, to find the minimum point.

After solving problem (\ref{eq:sub_amplitude_single_m2}) and getting the optimal amplitude $\alpha_{\mathrm{t},m}^\star$, we can obtain $\phi_{\mathrm{t/r},m}^\star$ as:
\begin{equation}
    \label{eq:single_opt_phi}
    \begin{aligned}
        \phi_{\mathrm{t},m}^\star = \sqrt{\alpha_{\mathrm{t},m}^\star}e^{\jmath\theta_{\mathrm{t},m}^\star}, 
        \phi_{\mathrm{r},m}^\star =  \sqrt{1 - \alpha_{\mathrm{t},m}^\star}e^{\jmath\theta_{\mathrm{r},m}^\star}, \forall m \in \mathcal{M}.
    \end{aligned}
\end{equation}
The details of the single-connected RIS beamforming design are summarized in Algorithm \ref{alg:Single}.

\subsection{Initialization}
Choosing an appropriate initialization of BD-RIS coefficients $\mathbf{\Phi}_\mathrm{t}$, $\mathbf{\Phi}_\mathrm{r}$ and transmit beamformer $\mathbf{W}$ is important for the proposed Algorithm \ref{alg:fp}. Unfortunately, it is not that easy to quickly find different and good initial values of $\mathbf{\Phi}_\mathrm{t}$, $\mathbf{\Phi}_\mathrm{r}$ satisfying different modes/connections. Therefore, we simply adopt CW-SC cases and assume initial $\mathbf{\Phi}_\mathrm{t}$, $\mathbf{\Phi}_\mathrm{r}$ are all diagonal matrices, i.e., $\mathbf{\Phi}_\mathrm{t/r} = \mathsf{diag}(\phi_{\mathrm{t/r},1}, \ldots, \phi_{\mathrm{t/r},M})$. Each non-zero element has constant amplitude $\frac{1}{\sqrt{2}}$ and random phase shift within the range $[0,2\pi)$.

Then with the above initial BD-RIS coefficients, we use a typical minimum mean square error (MMSE) precoder to initialize $\mathbf{W}$, which is given by
\begin{equation}
    \tilde{\mathbf{W}} = (\mathbf{G}^H\mathbf{\Phi}^H\mathbf{H}\mathbf{H}^H\mathbf{\Phi}\mathbf{G} + \sigma^2 \mathbf{I}_N)^{-1}\mathbf{G}^H\mathbf{\Phi}^H\mathbf{H},
\end{equation}
where $\mathbf{\Phi} = [\mathbf{\Phi}_\mathrm{t}^H, \mathbf{\Phi}_\mathrm{r}^H]^H$, $\mathbf{H} = \mathsf{blkdiag}(\mathbf{H}_\mathrm{t},\mathbf{H}_\mathrm{r})$ with $\mathbf{H}_\mathrm{t/r} = [\mathbf{h}_{\mathrm{t/r},1}, \ldots, \mathbf{h}_{\mathrm{t/r},K_\mathrm{t/r}}]$, $\sigma_{i,k} = \sigma$, $\forall i \in \{\mathrm{t,r}\}$, $\forall k \in \mathcal{K}_i$. Finally, an additional normalization is performed to satisfy the power constraint, i.e., $\mathbf{W} = \frac{\sqrt{P}\tilde{\mathbf{W}}}{\|\tilde{\mathbf{W}}\|_F}$.

\begin{figure}[!t]
    \centering
    \subfigure[Rayleigh fading]{
    {\label{fig:SR_Iter_64_rayleigh}}
    \includegraphics[height=1.45 in]{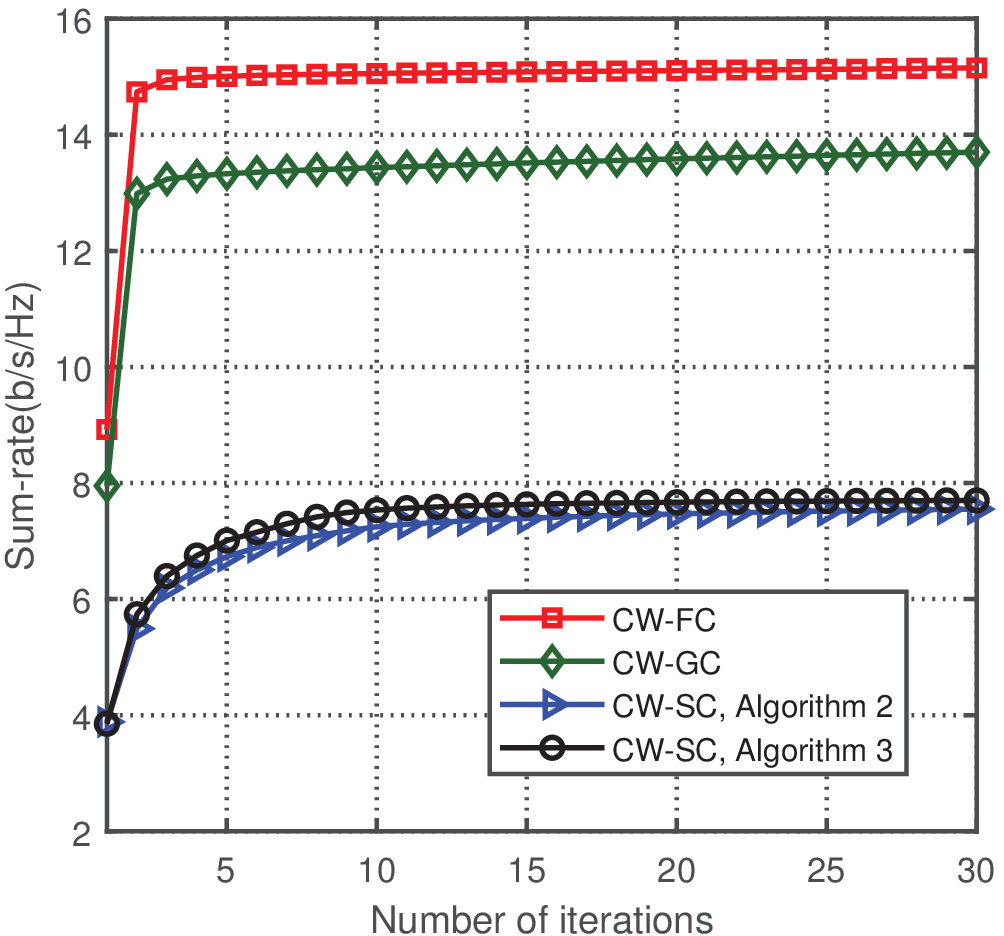}}
    \subfigure[Rician fading]{
    {\label{fig:SR_Iter_64_rician}}
    \includegraphics[height=1.45 in]{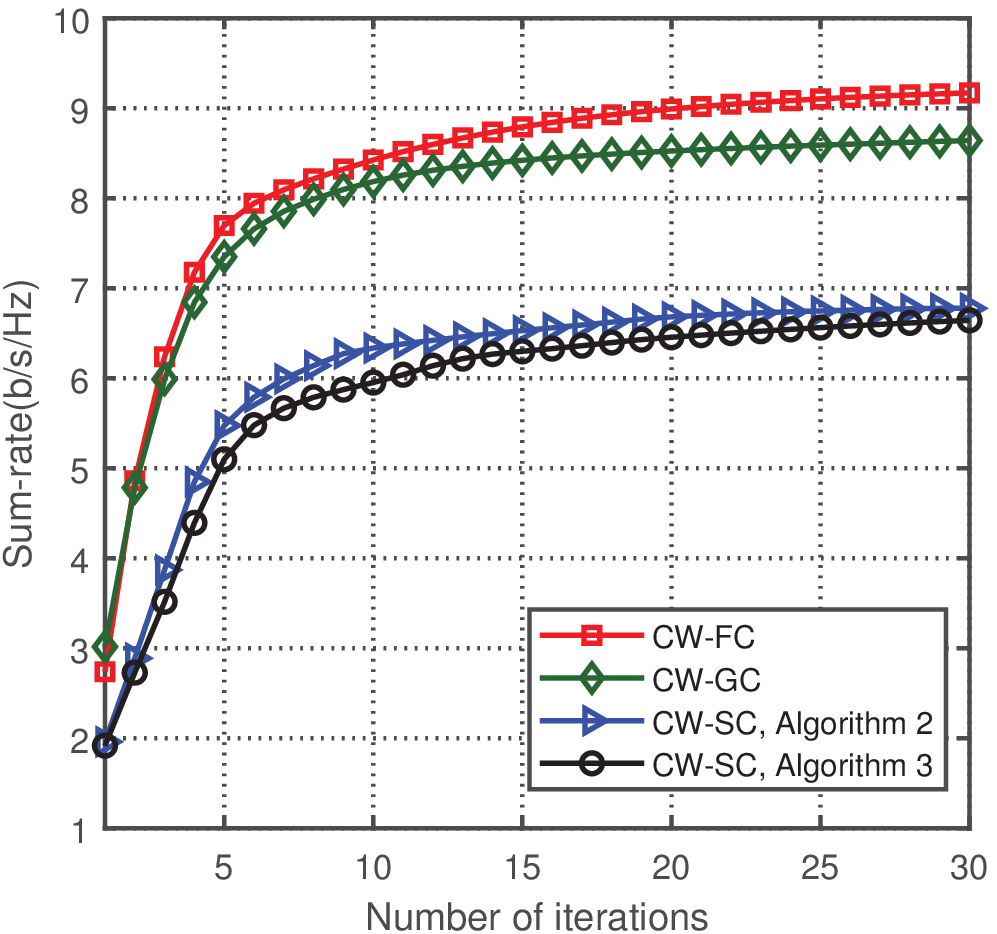}}
    \caption{Sum-rate versus the number of iterations ($P = 5$ dBm, $G = 8$, $N = K = 6$, $K_\mathrm{t} = K_\mathrm{r} = 3$, $M = 64$).}
    \label{fig:sr_vs_iter}
\end{figure}

\begin{table*}[t]
    \caption{The Circuit Topology Complexity and Overall Optimization Complexity for BD-RIS with Different Architectures}
    \centering
        \begin{tabular}{|c|c|c|c|c|c|}
            \hline
            Architecture & \makecell{Number of \\ Groups} & Group Size & \makecell{Number of \\ Non-zero Elements} & \makecell{Circuit Topology\\ Complexity} &Optimization Complexity\\
            \hline
            \multirow{2}*{CW-SC} & \multirow{2}*{$M$} & \multirow{2}*{1} & \multirow{2}*{$2M$} & \multirow{2}*{$3M$} & $\mathcal{O}\{I(K^2M^2 + I_\mathrm{bs}KN^3 + I_\mathrm{gc}I_\mathrm{cg}M)\}$ by Algorithm \ref{alg:MCG}\\
            \cline{6-6}
            \multirow{2}*{} & \multirow{2}*{} & \multirow{2}*{} & \multirow{2}*{} & \multirow{2}*{} &$\mathcal{O}\{I(K^2M^2 + I_\mathrm{bs}KN^3 + I_\mathrm{sc}M)\}$ by Algorithm \ref{alg:Single}\\
            \hline
            CW-GC & $G$ & $\bar{M}$ & $2G\bar{M}^2$ & $M(2\bar{M}+1)$ & $\mathcal{O}\{I(K^2M^2 + I_\mathrm{bs}KN^3 + I_\mathrm{gc}I_\mathrm{cg}G\bar{M}^3)\}$ \\
            \hline
            CW-FC & 1 & $M$ &$2M^2$ & $M(2M+1)$ &$\mathcal{O}\{I(K^2M^2 + I_\mathrm{bs}KN^3 + I_\mathrm{gc}I_\mathrm{cg}M^3)\}$\\
            \hline
        \end{tabular}\label{Table_2}
\end{table*}

\subsection{Convergence Analysis}
The convergence of the proposed Algorithm \ref{alg:fp} cannot be strictly proved, since the update of block $\{\mathbf{\Phi}_\mathrm{t}, \mathbf{\Phi}_\mathrm{r}\}$ has no guarantee of global optimum. Fortunately, since the update of the remaining blocks (steps 3-5 in Algorithm \ref{alg:fp}) is monotonous, the loss induced by updating block $\{\mathbf{\Phi}_\mathrm{t}, \mathbf{\Phi}_\mathrm{r}\}$ is negligible. 
Although we cannot provide a rigorous theoretical proof for the convergence of Algorithm \ref{alg:fp}, we evaluate the convergence performance by simulations. Simulation results for the proposed algorithms are shown in Fig. \ref{fig:sr_vs_iter}. In Fig. \ref{fig:sr_vs_iter}, schemes ``CW-FC'', ``CW-GC'', and ``CW-SC, Algorithm \ref{alg:MCG}'' are plotted based on Algorithms \ref{alg:MCG} and \ref{alg:fp}, while ``CW-SC, Algorithm \ref{alg:Single}'' is plotted based on Algorithms \ref{alg:Single} and \ref{alg:fp}. From Fig. \ref{fig:sr_vs_iter} we can observe that our proposed solutions always converge within limited iterations under different channel realizations, e.g., Rayleigh fading and Rician fading channels. These simulation results demonstrate the robustness of our proposed algorithms. Moreover, our proposed efficient solution for CW-SC BD-RIS, i.e. ``CW-SC, Algorithm \ref{alg:Single}'', has a similar performance to ``CW-SC, Algorithm \ref{alg:MCG}'', but with a lower complexity (which will be discussed in the following subsection).

\subsection{Complexity Analysis}

In this subsection, we provide a broad complexity analysis for algorithms proposed in the previous section.  As shown in Algorithm \ref{alg:fp}, four blocks are iteratively updated to find a convergent solution. In each iteration, updating blocks ${\bm \iota}$ and ${\bm \tau}$ require $\mathcal{O}\{K^2M^2\}$ operations; updating block $\mathbf{W}$ has a complexity of approximately $\mathcal{O}\{K(M^2 + I_\mathrm{bs}N^3)\}$, where $I_\mathrm{bs}$ is the number of iterations for bisection search. The complexity of the proposed two algorithms for designing block $\mathbf{\Phi}_\mathrm{t}$, $\mathbf{\Phi}_\mathrm{r}$ will be discussed as follows.

\subsubsection{The General Solution for CW-GC BD-RIS} 
As summarized in Algorithm \ref{alg:MCG}, the optimization of CW-GC BD-RIS is based on an iterative design. In each iteration, the objective is divided into $G$ sub-problems, each of which is solved by a manifold version of the CG method (steps 7-13 in Algorithm \ref{alg:MCG}) with complexity $\mathcal{O}\{I_\mathrm{cg}\bar{M}^3\}$, where $I_\mathrm{cg}$ denotes the number of iterations for the CG method.  
Therefore, calculating CW-GC BD-RIS requires $\mathcal{O}\{I_\mathrm{gc}I_\mathrm{cg}G\bar{M}^3\}$ operations, where $I_\mathrm{gc}$ denotes the number of iterations for Algorithm \ref{alg:MCG}.
The overall complexity is $\mathcal{O}\{I(K^2M^2 + I_\mathrm{bs}KN^3 + I_\mathrm{gc}I_\mathrm{cg}G\bar{M}^3)\}$, where $I$ denotes the number of iterations in Algorithm \ref{alg:fp}.
Since CW-FC and CW-SC BD-RIS are special cases of CW-GC ones, we can easily derive the corresponding complexity:

\begin{itemize}
    \item[C1:] For the CW-FC case, i.e., $G = 1$, there is no need to do iterations for BD-RIS design so that the complexity of updating BD-RIS beamformer is $\mathcal{O}\{I_\mathrm{cg}M^3\}$. The overall complexity is thus $\mathcal{O}\{I(K^2M^2 + I_\mathrm{bs}KN^3 + I_\mathrm{cg}M^3)\}$.
    \item[C2:] For the CW-SC case, i.e., $G = M$, optimizing BD-RIS requires  $\mathcal{O}\{I_\mathrm{gc}I_\mathrm{cg}M\}$. And the overall complexity is $\mathcal{O}\{I(K^2M^2 + I_\mathrm{bs}KN^3 + I_\mathrm{gc}I_\mathrm{cg}M)\}$.
\end{itemize}

\subsubsection{The Efficient Solution for CW-SC BD-RIS}
Updating $\{\mathbf{\Phi}_\mathrm{t}, \mathbf{\Phi}_\mathrm{r}\}$ by Algorithm \ref{alg:Single} requires $\mathcal{O}\{I_\mathrm{sc}M\}$ operations, where $I_\mathrm{sc}$ denotes the number of iterations. Thus, the complexity for joint transmit beamformer and CW-SC BD-RIS matrix design is $\mathcal{O}\{I(K^2M^2 + I_\mathrm{bs}KN^3 + I_\mathrm{sc}M)\}$.

To summarize, different architectures have different circuit topologies, which require different numbers of reconfigurable impedance components. Therefore, mathematically $\mathbf{\Phi}_\mathrm{t}$ and $\mathbf{\Phi}_\mathrm{r}$ have different numbers of non-zero elements related to the number of cells $M$ and groups $G$, yielding different complexity for beamforming design. To provide a clear comparison, we summarize the circuit topology complexity and optimization complexity for different architectures in Table \ref{Table_2}, showing that the performance improvement (as illustrated in the next section) of cell-wise group/fully-connected BD-RIS is at the expense of higher circuit topology complexity and optimization complexity.

\section{Performance Evaluation}

\begin{figure}[!t]
    \centering
    \includegraphics[height=1.3 in]{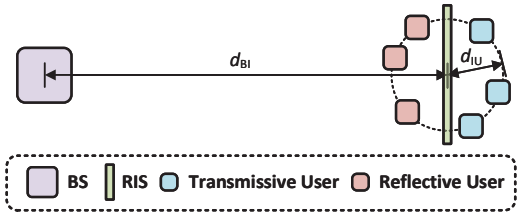}
    \caption{An illustration of the relative position among the BS, RIS, and users.}\label{fig:sim_set} 
\end{figure}

In this section, we present simulation results to demonstrate the performance of the BD-RIS-aided MU-MISO system when the BD-RIS has nine different modes/architectures. 
We model the channels from the BS to the RIS and from the RIS to users as a combination of small-scale and large-scale fading in accordance with existing RIS works \cite{RZhang,JXu,yildirim2020modeling}. Specifically, the distance-dependent pathloss model $PL_i = \zeta_0(d_i/d_0)^{-\varepsilon_i}$, $\forall i \in \{\mathrm{BI,IU}\}$ \cite{RZhang,JXu} accounts for the large-scale fading with $\zeta_0$ referring to the signal attenuation at a reference distance $d_0$, $d_\mathrm{BI}$ and $d_\mathrm{IU}$ referring to the distance between the BS and the RIS, and between the RIS and users, and $\varepsilon_i$, $\forall i \in \{\mathrm{BI,IU}\}$ referring to the path loss exponent. Regarding the small-scale fading, we use both Rayleigh fading and Rician fading models \cite{RZhang,JXu,yildirim2020modeling} with
$\kappa_i$, $\forall i \in \{\mathrm{BI,IU}\}$ referring to the Rician factor. In the following simulations, we fix the Rician factor for Rician fading channels to $\kappa_\mathrm{BI} = \kappa_\mathrm{IU} = 5$ dB for BS-RIS and RIS-user links.
The signal attenuation is set as $\zeta_0 = -30$ dB at a reference distance $d_0 = 1$ m for all channels. The path loss exponent of BS-RIS and RIS-user channels are all set as $\varepsilon_\mathrm{BI} = \varepsilon_\mathrm{IU} = 2.2$.
The noise power at each user is set as $\sigma_{k}^2 = -80$ dBm, $\forall i \in \{\mathrm{t,r}\}$, $\forall k \in \mathcal{K}_i$. The relative position among the BS, BD-RIS, and users is shown in Fig. \ref{fig:sim_set}, where the distance between the BS and the RIS is set as $d_\mathrm{BI} = 50$ m, and $K$ users are randomly located close to the RIS with the same distance $d_\mathrm{IU} = 2.5$ m.

\begin{figure}[!t]
    \centering
    \subfigure[Rayleigh fading]{
    \includegraphics[width=3 in]{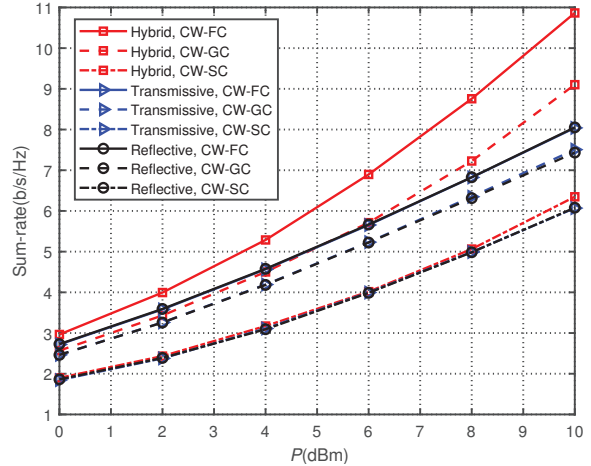}}
    \subfigure[Rician fading]{
    \includegraphics[width=3 in]{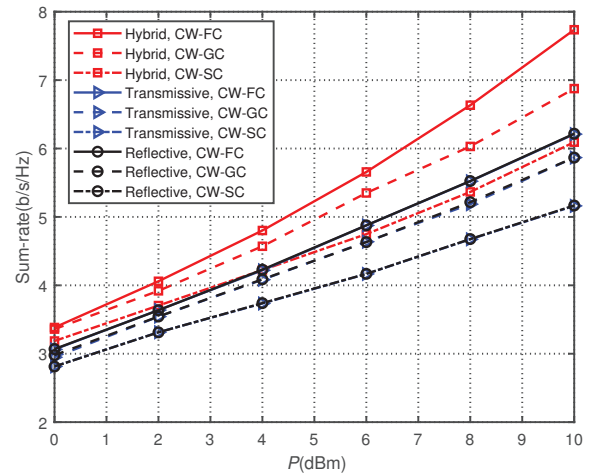}}
    \caption{Sum-rate versus transmit power $P$ ($N = K = 4$, $K_\mathrm{t} = K_\mathrm{r} = 2$, $M = 32$, $G = 8$).}\vspace{-0.3 cm}
    \label{fig:sr_vs_p}
\end{figure}

\begin{figure}[!t]
    \centering
    \subfigure[Rayleigh fading]{\includegraphics[width=3 in]{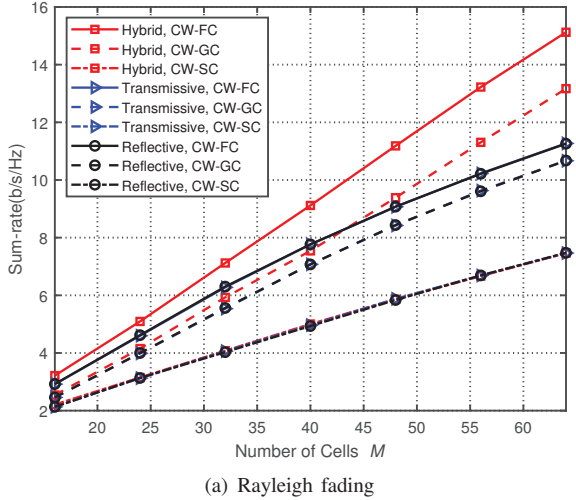}}
    \subfigure[Rician fading]{\includegraphics[width=3 in]{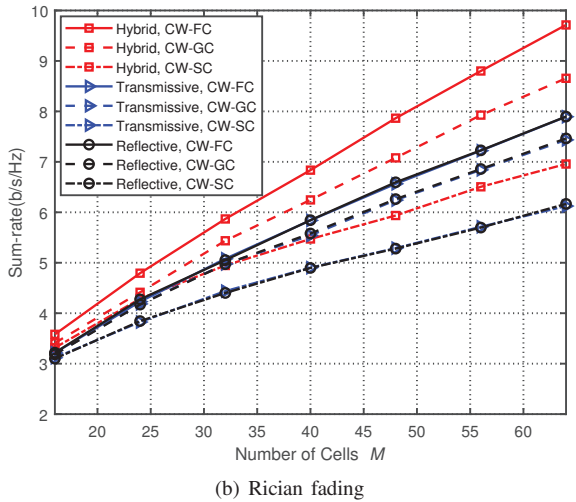}}
    \caption{Sum-rate versus the number of RIS cells $M$ ($N = K = 6$, $K_\mathrm{t} = K_\mathrm{r} = 3$, $G = 8$, $P = 5$ dBm).}
    \label{fig:sr_vs_m}\vspace{-0.3 cm}
\end{figure}

Now we examine the sum-rate performance of the BD-RIS-aided MU-MISO system when BD-RIS has different modes\footnote{We can easily obtain the RIS beamformer under the reflective (transmissive) mode based on the proposed algorithms by removing transmissive (reflective) users and forcing $\mathbf{\Phi}_\mathrm{t}$ ($\mathbf{\Phi}_\mathrm{r}$) to zero.} and architectures\footnote{In the following simulations, CW-SC BD-RIS beamformer is designed by Algorithm \ref{alg:Single}.}.
Fig. \ref{fig:sr_vs_p} shows the sum-rate performance versus the transmit power. 
From Fig. \ref{fig:sr_vs_p} we can obtain the following conclusions: \textit{i)} With the same architecture, the ``hybrid'' scheme can outperform other modes. For example, for Rician fading channels, the ``hybrid, CW-FC'' scheme can achieve around 20\% higher sum-rate than ``Transmissive/Reflective, CW-FC'' schemes. This is because the hybrid BD-RIS can fully utilize the multiuser diversity, while part of users are blocked under other modes. Therefore, the ``hybrid'' scheme can realize a full-dimensional service coverage, which indicates that the scope of application for hybrid RISs can be much more flexible than ``transmissive/reflective'' ones. \textit{ii)} Under the same mode, the ``full'' scheme always achieves the best sum-rate performance. 
For example, for Rayleigh fading channels, the sum-rate of ``hybrid, CW-FC'' and ``hybrid, CW-GC'' schemes is around 75\% and 37\% higher than that of the ``hybrid, CW-SC'' scheme.  
\textit{iii)} When the transmit power is relatively small (the sum-rate performance is more related to the BD-RIS matrix than to the BS precoder), with the same inter-cell architecture, the performance gap between the hybrid RIS and transmissive/reflective RIS for Rician fading channels is larger than that for Rayleigh fading channels. It is because in Rician fading channels with strong LoS components, two separate beams of the hybrid BD-RIS can be ``sharper'' to finely direct to users from different sides.    
Therefore, hybrid RISs are more suitable in Rician fading environments. 
\textit{iv)} When RISs are under the same mode (in low transmit power case), the performance gap between CW-FC/GC RISs and CW-SC ones for Rayleigh fading channels is larger than that for Rician fading channels. 
This fact shows that the advantage of CW-FC/GC RIS is more prominent in Rayleigh fading propagations, therefore confirming results in \cite{SShen}.

Then in Fig. \ref{fig:sr_vs_m}, we plot sum-rate as a function of the number of cells $M$. It can be observed from Fig. \ref{fig:sr_vs_m} that the sum-rate for all schemes grows with the increase of $M$. 
More importantly, the slope of the sum-rate-versus-$M$ curve for CW-FC/GC BD-RIS is larger than that for CW-SC BD-RIS. This phenomenon can be explained by Table \ref{Table_2}, which demonstrates that the number of non-zero elements grows in a quadratic manner with increasing $M$ for CW-FC/GC BD-RIS, but grows linearly for CW-SC case: The more the number of non-zero elements, the higher the beam control flexibility.

\begin{figure}[!t]
    \centering
    \subfigure[Rayleigh fading ($K_\mathrm{r} = 2$)]{
    {\label{fig:SR_Kt_rayleigh}}
    \includegraphics[height=1.3 in]{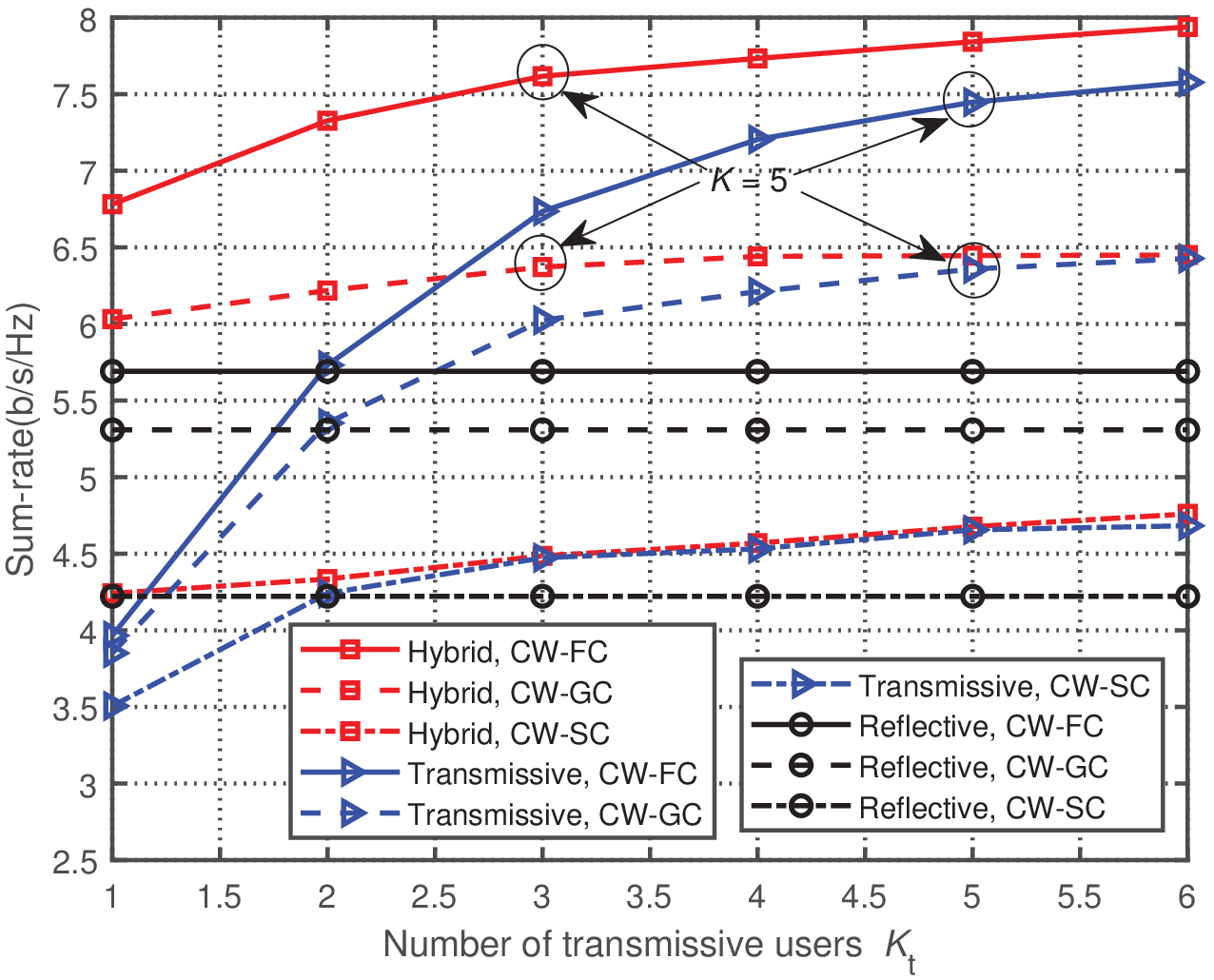}}
    \subfigure[Rayleigh fading ($K_\mathrm{t} = 2$)]{
    {\label{fig:SR_Kr_rayleigh}}
    \includegraphics[height=1.3 in]{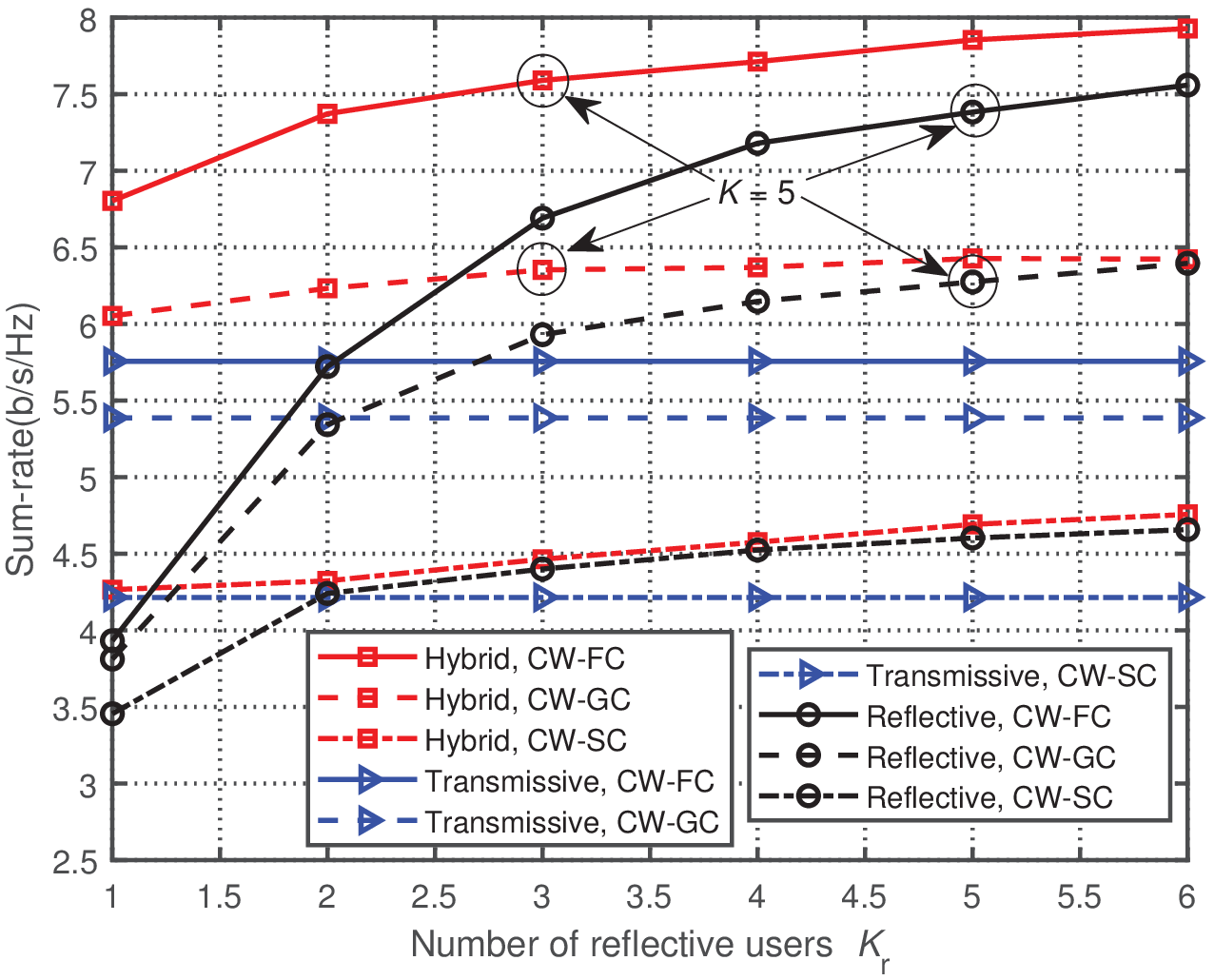}}
    \subfigure[Rician fading ($K_\mathrm{r} = 2$)]{
    {\label{fig:SR_Kt_rician}}
    \includegraphics[height=1.3 in]{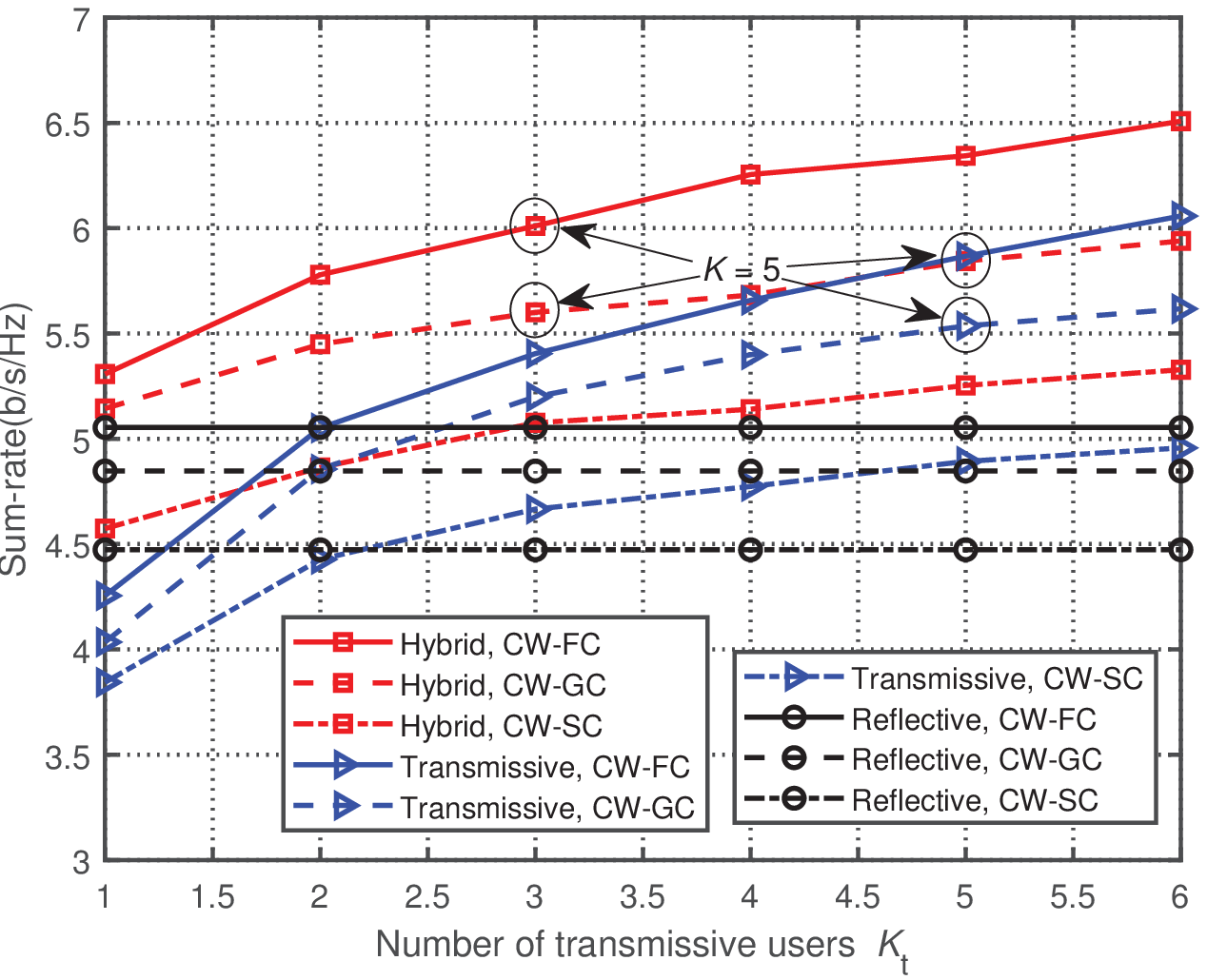}}
    \subfigure[Rician fading ($K_\mathrm{t} = 2$)]{
    {\label{fig:SR_Kr_rician}}
    \includegraphics[height=1.3 in]{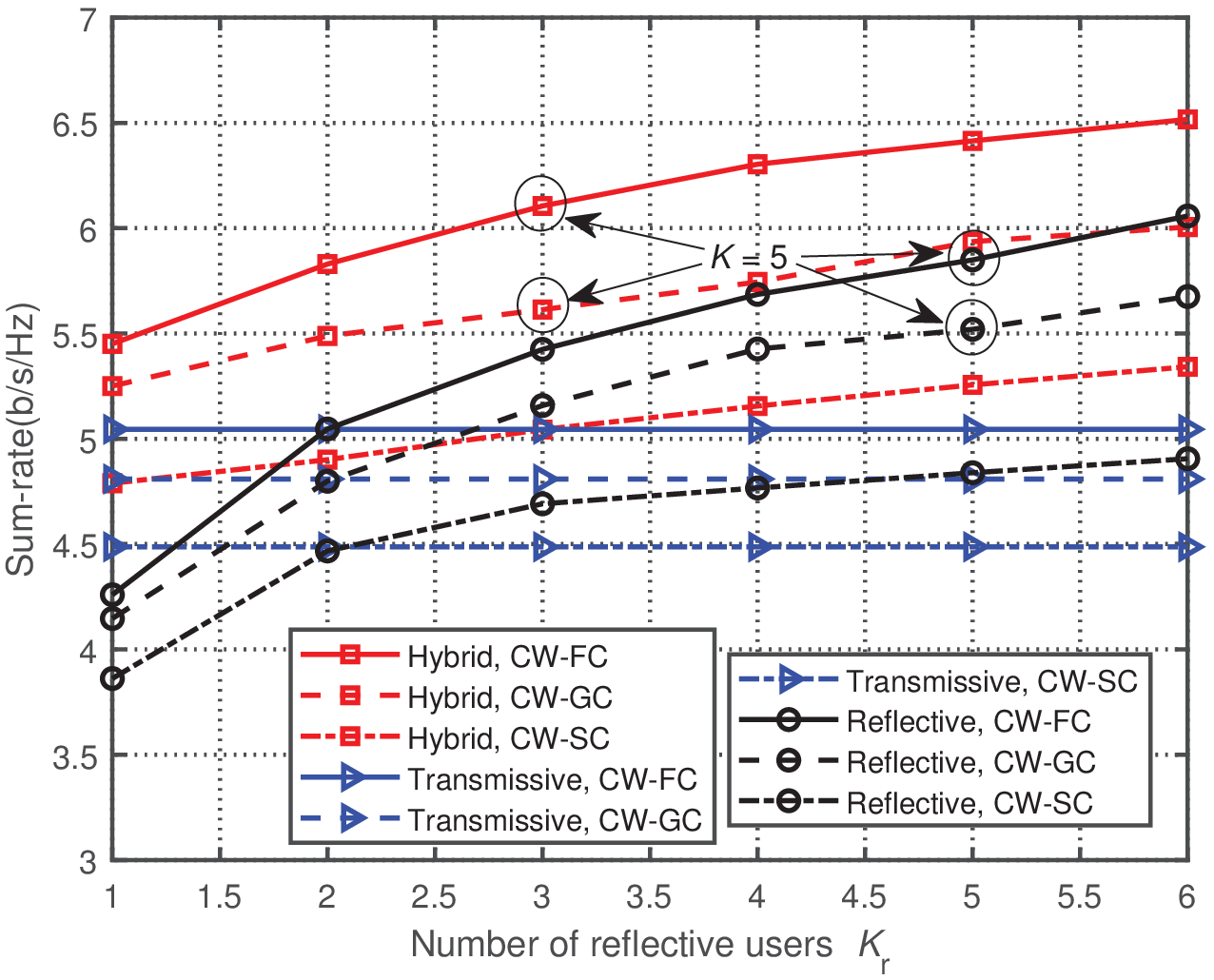}}
    \caption{(a), (c): Sum-rate versus the number of transmissive users; (b), (d): Sum-rate versus the number of reflective users ($M = 32$, $N = 8$, $P = 5$ dBm, $G = 8$).}\vspace{-0.3 cm}
\end{figure}

Finally, in Figs. \ref{fig:SR_Kt_rayleigh}, \ref{fig:SR_Kt_rician} and Figs. \ref{fig:SR_Kr_rayleigh}, \ref{fig:SR_Kr_rician}, we plot sum-rate versus the number of transmissive and reflective users under different channel fadings. From Figs. \ref{fig:SR_Kt_rayleigh}-\ref{fig:SR_Kr_rician} we have the following observations: \textit{i)} With the same architecture and same numbers of transmissive/reflective users, the ``hybrid'' scheme can always outperform its competitors. \textit{ii)} With the growth of the number of transmissive/reflective users, the performance gap between the ``hybrid'' scheme and the ``transmissive/reflective'' one becomes smaller. \textit{iii)} When the number of total users the BS can simultaneously serve is fixed, e.g., circled points in Figs. \ref{fig:SR_Kt_rayleigh}-\ref{fig:SR_Kr_rician}, the ``hybrid'' scheme still achieves better performance than ``reflective/transmissive'' schemes. This phenomenon can be explained as follows. For the ``hybrid'' scheme, there are two different beams directing to users from two sides, while for the ``reflective/transmissive'' scheme, all the users share the same beam. More specifically, the ``reflective/transmissive'' scheme can be regarded as special cases of the ``hybrid'' scheme (according to the mathematical constraint summarized in Table I), which can realize more flexible beam control.

\section{Conclusion and Future Work}
\label{sc:Conclusion}

In this paper, we analyze and propose a general BD-RIS-aided communication model unifying three modes, namely, reflective, transmissive, and hybrid, and three architectures, namely, CW-SC, CW-GC, and CW-FC architectures, totally nine cases. Particularly, we first prove that the STAR-RIS (hybrid CW-SC RIS) is a special case of our proposed model. Then we go beyond STAR-RIS and propose more flexible CW-GC/FC architectures. 

With the proposed model, we consider the joint transmit precoder and BD-RIS matrix design to maximize the sum-rate for an RIS-aided MU-MISO system. We first transform the original problem into a multi-block optimization based on fractional programming theory. 
For the design of BD-RIS block, we propose a general algorithm, which is suitable for different architectures, and an efficient algorithm specifically for CW-SC BD-RIS. 

Finally, we provide a comprehensive comparison for BD-RIS with nine modes/architectures. 
Simulation results show that under specific parameter settings and channel conditions, CW-FC and CW-GC hybrid BD-RISs can achieve around 75\% and 37\% higher sum-rate performance than CW-SC hybrid BD-RISs. Meanwhile, using CW-FC hybrid BD-RISs can improve the sum-rate performance by around 20\% compared with using CW-FC transmissive/reflective ones.

Based on the general and unified BD-RIS model, there are many issues worth being studied for future research, which include, but are not limited to the following aspects:
\subsubsection{Wideband Modeling} Currently only a few results consider the wideband modeling of RIS \cite{WCai,HLi,WYang} restricted to diagonal phase shift matrices. In the future, it is interesting and important to extend our BD-RIS model to wideband cases and reconsider the beamforming design. 
\subsubsection{Interplay with Other Techniques} Given the advantages on performance enhancement of the proposed CW-GC/FC BD-RIS, it is interesting to explore the applications of such model by considering the integration of BD-RIS with other techniques, such as rate-splitting multiple access (RSMA), WPT, SWIPT, integrated sensing and communication (ISAC).
\subsubsection{Prototyping} How to physically realize the proposed BD-RIS is a very important open problem. There have been some works related to the hardware implementation of STAR-RIS \cite{guo2019transmission,bao2021programmable,song2021switchable}, which can be leveraged to facilitate the implementation of the proposed BD-RIS. In the future, it will be meaningful to consider the prototyping of the more general CW-GC/FC BD-RIS to experimentally verify its advantages on performance enhancement.

\begin{appendix}[Proof of Convexity of Objective (\ref{eq:obj_amplitude2})]
    We first abstract objective as a real-value function $\bar{z}(x) = ax^2 + bx + c\sqrt{1 - x^2}$ with one-to-one correlations, i.e., $x = \sqrt{\alpha_{\mathrm{t},m}} \in (0,1)$, $a = \upsilon_m \in \mathbb{R}$, $b = -2|\chi_{\mathrm{t},m}| < 0$, $c = -2|\chi_{\mathrm{r},m}| < 0$. Then the proof is given as follows:
    \begin{proof}
        Calculate first- and second-order derivatives of $\bar{z}(x)$ as $\triangledown  \bar{z}(x) = 2ax + b - \frac{cx}{\sqrt{1-x^2}}$ and $\triangledown^2 \bar{z}(x) = 2a - \frac{c}{(1 - x^2)^{\frac{3}{2}}}$.
        Then we discuss $\triangledown \bar{z}(x)$ under the following two conditions:
        \begin{enumerate}[C1:]
            \item When $a \ge 0$, it is obvious that $\triangledown^2\bar{z}(x) > 0$ so that $\triangledown \bar{z}(x)$ is a monotonically increasing function with $x \in (0,1)$.
            Based on the above discussion, we have $\lim_{x\rightarrow 0} \triangledown \bar{z}(x) = b < 0$, and $\lim_{x\rightarrow 1} \triangledown \bar{z}(x) = +\infty > 0$, 
            which indicate that $\bar{z}(x)$ is first monotonically decreasing and then monotonically increasing for $x \in (0,1)$.
            \item When $a < 0$, calculate the third-order derivative of $\bar{z}(x)$ as $\triangledown^3 \bar{z}(x) = -3cx(1-x^2)^{-\frac{5}{2}}$,
            which is always larger than zero when $x \in (0,1)$. Thus, $\triangledown^2\bar{z}(x)$ is monotonically increasing with $x \in (0,1)$. Then we calculate
                $\lim_{x\rightarrow 0} \triangledown^2 \bar{z}(x) = 2a - c$, whose value has the following two conditions:
            \begin{enumerate}[C2.1:]
                \item When $2a-c \ge 0$, $\triangledown^2 \bar{z}(x) > 0$ for $x \in (0,1)$. Thus, $\triangledown \bar{z}(x)$ is also monotonically increasing within the range $x \in (0,1)$ and the same conclusion as C1 can be obtained.
                \item When $2a-c < 0$, examine 
                    $\lim_{x\rightarrow 1} \triangledown^2 \bar{z}(x) = +\infty > 0$,
                which proves that $\triangledown \bar{z}(x)$ is first monotonically decreasing and then increasing within the range $x \in (0,1)$. Therefore, we can derive that $\bar{z}(x)$ has the same trend as $ \triangledown \bar{z}(x)$ for $x \in (0,1)$.
            \end{enumerate}      
        \end{enumerate}
        To summarize, $\bar{z}(x)$ is a convex function with only one minimum point for $x \in (0,1)$, which completes the proof.
    \end{proof}
\end{appendix}

\bibliographystyle{IEEEtran}
\bibliography{references}
\end{document}